\documentclass[sn-mathphys-num]{sn-jnl}% Math and Physical Sciences Numbered Reference Style 
\usepackage{graphicx}%
\usepackage{multirow}%
\usepackage{amsmath,amssymb,amsfonts}%
\usepackage{amsthm}%
\usepackage{mathrsfs}%
\usepackage[title]{appendix}%
\usepackage{xcolor}%
\usepackage{textcomp}%
\usepackage{manyfoot}%
\usepackage{booktabs}%
\usepackage{algorithm}%
\usepackage{algorithmicx}%
\usepackage{algpseudocode}%
\usepackage{listings}%

\usepackage[english]{babel}
\usepackage{blindtext}
\usepackage[utf8x]{inputenc} 
\usepackage{mathtools}
\usepackage{subcaption}
\usepackage{float}
\usepackage{xcolor}
\usepackage{ragged2e}
\newcommand{\eqn}[2]{\begin{equation}\label{#1}#2 \end{equation}}
\newcommand{\Mpl}{M_{\rm pl}}

\theoremstyle{thmstyleone}%
\theoremstyle{thmstyletwo}%

\theoremstyle{thmstylethree}%

\begin{document}

\title[Article Title]{An \'Etude on the Regularization and Renormalization of Divergences in Primordial Observables}

\author[1]{\fnm{Anna} \sur{Negro}}\email{negro@lorentz.leidenuniv.nl}

\author[1]{\fnm{Subodh P.} \sur{Patil}}\email{patil@lorentz.leidenuniv.nl}

\affil[1]{\orgdiv{Instituut-Lorentz for Theoretical Physics}, \orgname{Leiden University}, \orgaddress{\street{Niels Bohrweg}, \city{Leiden}, \postcode{2333 CA}, \country{The Netherlands}}}

\abstract{Many cosmological observables derive from primordial vacuum fluctuations evolved to late times. These observables represent statistical draws from some underlying quantum or statistical field theoretic framework where infinities arise and require regularization. After subtraction, renormalization conditions must be imposed by measurements at some scale, mindful of scheme and background dependence. We review this process on backgrounds that transition from finite duration inflation to radiation domination, and show how in spite of the ubiquity of scaleless integrals, ultra-violet (UV) divergences can still be meaningfully extracted from quantities that nominally vanish when dimensionally regularized. In this way, one can contextualize calculations with hard cutoffs, distinguishing between UV and infra-red (IR) scales corresponding to the beginning and end of inflation from UV and IR scales corresponding the unknown completion of the theory and its observables. This distinction has significance as observable quantities cannot depend on the latter although they will certainly depend on the former. One can also explicitly show the scheme independence of the coefficients of UV divergent logarithms. Furthermore, certain IR divergences are shown to be an artifact of the de Sitter limit and are cured for finite duration inflation. For gravitational wave observables, we stress the need to regularize stress tensors that do not presume a prior scale separation in their definition (as with the standard Isaacson form), deriving an improved stress tensor fit to purpose. We conclude by highlighting the inextricable connection between inferring $N_{\rm eff}$ bounds from vacuum tensor perturbations and the process of background renormalization.}


\maketitle

\section{Introductory remarks}

Infinities often arise when computing physical observables in any quantum or statistical field theoretic framework. A well defined prescription to regularize and renormalize these divergences has been developed on flat backgrounds\footnote{Among the many references one could consult on this topic, the treatments in \cite{Kleinert:2001ax} and \cite{Burgess:2020tbq} in the context of renormalizable, and effective field theories, respectively, provide comprehensive and pedagogical introductions.}, and a transposition of this formalism to curved backgrounds is also possible, albeit with numerous caveats and subtleties that one must be mindful of depending on the context \cite{Birrell:1982ix, Parker:2009uva, Mukhanov:2007zz}\footnote{A number of practical corollaries follow from the idealization of a Minkowski background that no longer apply on curved backgrounds. These include the ability to define one particle states as irreducible unitary representations of the background symmetry group \cite{Weinberg:1995mt}, the existence of a global timelike Killing vector through which positive and negative frequency solutions can be defined \cite{Birrell:1982ix, Parker:2009uva} and energy scales cleanly identified to impose a separation of scales \cite{Burgess:2020tbq}, in addition to the existence of in and out states and asymptotic flatness that premise the existence of an S-matrix \cite{Wald:1979wt}.}.

Divergences can generically be classed into two categories with distinct physical interpretations. Short distance, or ultra-violet (UV) divergences are unambiguously interpreted as an artifact of admitting arbitrarily high energies or momenta in any given calculational setup as a mathematical idealization. Once these infinities have been appropriately subtracted, causality and locality conspire through decoupling \cite{Symanzik:1970rt, Appelquist:1974tg} to limit the consequences of this idealization to a handful of relevant and marginal operators in a parametrically controlled derivative expansion (see e.g. \cite{Kleinert:2001ax, Burgess:2020tbq, Manohar:2018aog}). 

Long wavelength, or infra-red (IR) divergences on the other hand require additional care in their interpretation, as their implications can range from the harmless to the severe. At the very least, they indicate that one has additional work to do in order to arrive at a physical quantity. For instance, IR divergences often feature in intermediate expressions in theories with gapless excitations, and are rendered benign through cancellations when computing physical observables (as is the case in QED\footnote{Where IR divergences due to arbitrarily large numbers of soft quanta are compensated by phase space factors in certain scattering processes \cite{Bloch:1937pw}.}). IR divergences could also be signaling the breakdown of a particular perturbative scheme, which can nevertheless be tamed through resummation, as is the case when one studies thermalization in an out of equilibrium context \cite{Calzetta:2008iqa, Berges:2004yj, Maciejko}, or in QCD scattering via the factorization properties of its soft and collinear limits \cite{Kunszt:1996cy, Luisoni:2015xha}. In their most problematic incarnation, IR divergences are a harbinger of the instability of the putative background one is attempting perturbation theory around. Precisely which of the latter two possibilities are signaled by certain IR divergences encountered on de Sitter (dS) or quasi de Sitter backgrounds remains an open question (see e.g. \cite{Ford:1984hs, Mottola:1985qt, Antoniadis:1986sb, Dolgov:1994ra, Anderson:2009ci, Higuchi:2008tn, Burgess:2010dd, Giddings:2010nc, Onemli:2004mb, Kahya:2006hc, Polyakov:2012uc, Akhmedov:2012dn, Akhmedov:2013xka, Anderson:2013ila, Green:2020txs}). 

In operational terms, a class of IR divergences in cosmology trace to complications that arise on time dependent backgrounds when standard background field quantization methods can no longer be taken for granted. Large quantum corrections to the background could invalidate any purported classical background/fluctuation split, which is a limiting case of the general complication of doing perturbation theory around a time evolving background that must satisfy the tadpole condition at any given order in $\hbar$. Although addressing these issues may seem to necessitate the full regalia of non-equilibrium techniques, notable efforts within the context of stochastic inflation and its background statistical field theory limit can be found in \cite{Starobinsky:1986fx, Bardeen:1986iq, Rey:1986zk, Goncharov:1987ir, Nambu:1988je, Nakao:1988yi, Nambu:1989uf, Mollerach:1990zf, Salopek:1990re, Linde:1993xx, Starobinsky:1994bd, Finelli:2008zg, Enqvist:2008kt, Finelli:2010sh, PerreaultLevasseur:2013kfq, Vennin:2015hra, Assadullahi:2016gkk, Ezquiaga:2019ftu, Ando:2020fjm, Ballesteros:2020sre, Tada:2021zzj, Cruces:2022imf}, as well as within the dynamical renormalization group and open effective field theory approaches \cite{Boyanovsky:2003ui, Burgess:2009bs, Burgess:2015ajz}. 

Nevertheless, in situations where one can presume the validity of background field quantization, dealing with divergences in a variety of cosmological applications is a well understood prescription\footnote{See for instance \cite{Bartolo:2007ti, Seery:2010kh} for a review on how to extract physically meaningful answers for well defined observables in spite of the nominal presence of IR divergences in inflationary cosmology.}. So much so, that it is often taken for granted and elided in many practical applications in the literature where the ability to meaningfully extract physically observables quantities is taken as a given once divergences have purportedly been tamed. However, as the adage goes, just because something is infinite, does not mean it is zero\footnote{\label{quote}Unattributed quote recounted in Ch. 1.3 of \cite{Weinberg:1995mt}.}. The purpose of this article is to review the details of subtracting and renormalizing divergent quantities on backgrounds that should be of practical interest to all cosmologists -- one that transitions from an arbitrary but finite duration of inflation through to a phase of radiation domination. Doing so throws up qualitative differences and novelties relative to the analogous procedure on backgrounds corresponding to a single epoch (whether the inflationary, radiation, or matter dominated backgrounds) that warrants a discussion even for the simplest possible scenario of non-interacting, minimally coupled test scalar fields.

Foremost, we note the ubiquity of scaleless integrals for sufficiently light (though not necessarily massless) fields even when one transitions between epochs at a particular energy scale. Nevertheless, despite the fact that scaleless integrals nominally vanish in mass independent regularization schemes, it is still possible to cleanly separate UV and IR divergent contributions (using techniques borrowed from matching calculations in non-relativistic QCD and QED \cite{Burgess:2020tbq, Manohar:1997qy}), and compare these to the contributions one would have calculated in mass dependent regularization schemes, such as through imposing hard cutoffs in physical momenta\footnote{The moral of which seemingly inverts the previous adage: just because something is zero, doesn't mean it isn't infinite.}. This is useful, as UV and IR divergences have distinct physical interpretations -- whereas UV divergences are to be subtracted with local counterterms, IR divergences cancel among themselves once physical observables are computed\footnote{With the caveat that well defined observables can be identified, which presume that some of the IR divergences in question are not signaling the invalidity of the particular background quantization.}. Moreover, even as one finds the expected scheme dependence for the coefficients of nominal power law divergences, one finds identical coefficients for UV divergent logarithms whether one dimensionally regularizes, point splits, or uses hard cutoffs in physical momenta.

Furthermore, one can demonstrate that certain IR divergences encountered in inflationary cosmology are an artifact of the idealization of a past infinite dS geometry and are absent in computations on backgrounds that transition in and out of inflation, whereas unsurprisingly, UV divergences persist. Specifically, we show how the end of inflation does not regulate UV divergences, but instead parameterizes the coefficients of divergent quantities through the ratio of the scales corresponding to the beginning and end of inflation and must still be subtracted in the prescribed way. The implications of this for vacuum contributions to measurable quantities with blue spectra is discussed, as one must be cautious against including contributions that could be nothing more than the Fourier transform of a UV divergence that is to be subtracted on the way to computing observable quantities.

We conclude with a discussion on following through on renormalizing divergences after having subtracted them in the context of calculations that attempt to compute quantum corrections to a given (tree level) background in some form. In simple terms, it is to be stressed that all physical observations are necessarily of fully dressed quantities, and the putative classical background one performs background field quantization around is only to be viewed as a calculational fiction that serves as bootstrap into computing physical observables. Calculations of quantum corrections tell us nothing of their absolute values. Once renormalization conditions are fixed by measurements at a particular scale, all one can compute is how these quantities change with scale. We discuss the implications of this point for the interpretation of $N_{\rm eff}$ bounds from vacuum tensor perturbations in the early universe, which can be boiled down to attempting to calculate a corrected background (i.e. shifted tadpole condition) from an appropriately derived stress tensor for gravitational waves. A definition of the latter that is fit for the purposes of regularization forces us to work with a form that does not assume a prior (time dependent) scale separation in its definition, as is the case with the commonly used Isaacson form. We retrace the derivation of the stress tensor for gravitational waves and derive an improved version suitable for regularization. We comment in closing how can one in principle compute the stress tensor for gravitational waves in a fully covariant formalism, bypassing entirely the need to compute it via correlation functions through Hadamard regularization techniques.

\subsection{Outline and scope}

The outline of this article is as follows, we first fix conventions and recap how UV and IR divergences arise in computations of standard cosmological observables, stressing that tracers corresponding to energy densities must be extractable from regularized stress tensors or derivatives of two point functions and their convolutions. For sufficiently light degrees of freedom, we show the ubiquity of scaleless integrals, taking care to address the order of various limits when one looks to extract divergences. Taking a cue from non-relativistic QED and QCD, we then show how to split scaleless integrals into UV and IR contributions that otherwise cancel in dimensional regularization. Focusing only on the UV log-divergent parts, one can make sense of calculations in the literature that impose hard cutoffs in physical momenta, finding identical coefficients for the UV divergent logs no matter the scheme. We then consider the process of regularization on backgrounds that transition in and out of inflation to radiation domination. Although scales corresponding to the beginning and end of inflation can be shown to merely parameterize rather than regulate UV divergences, one finds that nominally IR divergent contributions on a dS background are cured in the examples we consider\footnote{Whether this generalizes to a wider class of IR divergences, especially when higher point interactions are included is an important question that demands to be followed up on.}. In attempting to repeat this process for divergences associated with vacuum tensor perturbations, one immediately encounters the need to work with a form of the stress tensor that does not assume a prior scale separation in its construction if one is to integrate over all momenta in intermediate steps. We derive an improved stress tensor for gravitational waves, highlighting its advantages relative to the Isaacson form for applications to primordial observables, and address the relevance of our observations for extracting various cosmological bounds, before offering our concluding remarks. Various details are deferred to the appendices. 

This article is intended as a practical tour for the regularization and renormalization of nominally divergent quantities on backgrounds of cosmological interest using well established techniques, although applied to a context where following through on the details is informative to the point of novelty. For this reason, we focus on the simplest possible applications one could think of given the richness of the problems already encountered there -- that of minimally coupled non-interacting test scalar fields, and vacuum tensor perturbations sourced by the background expansion.  Moreover, we work in a formalism that should be familiar to most practicing cosmologists: two-point functions, power spectra, and  spectral densities, all constructed through the intermediary of mode functions derived on a Friedmann-Robertson-Lema\^itre-Walker (FRLW) slicing, as opposed to working in a fully covariant formalism (see e.g. \cite{Birrell:1982ix, Parker:2009uva} and references therein). The reason for this is as practical as it is editorial -- although the classic references exert a great deal of effort in the study of the regularization of divergences, they are notably sparse in detail when it comes to the process of following through on renormalization though imposing renormalization conditions\footnote{This is perhaps a historical corollary of the fact that many of the foundational references were written in a time where measurements of the observables they were considering (necessary for the imposition of renormalization conditions) seemed like a distant prospect, which is no longer the case.}, which are typically imposed on observables that convolve Fourier space quantities. In the interests of accessibility for the target audience, full covariance is sacrificed for clarity.

\section{Preliminaries -- regularization of stress tensors and correlation functions}
All physical observations consist of some sort of energy or momentum transfer between propagating degrees of freedom and some form of detector or tracer, whether directly or indirectly. Although correlation functions serve as a convenient calculational intermediary between any given (effective) description valid during the primordial epoch and observables at late times, they are not necessarily directly observable by themselves. They can, however, be related to observations at later times by acting upon them with the appropriate derivatives to extract energy and momentum, and subsequently convolving them with transfer functions that encapsulate how they are processed by the intervening cosmological evolution (see e.g. \cite{Chluba:2015bqa} for a review). Correlation functions are typically computed in Fourier space, whereas the observations they relate to are made at some fixed temporal and spatial location and involve the coincident limits of bilinear or higher point functions, necessitating subtraction of UV divergences associated with this limit. Well defined coincident limits also implicitly depend on the IR behavior of correlation functions in that any divergences encountered in the latter must cancel between all contributions in convolution with the transfer function for well defined physical observables in any self-consistent calculational setup. In what follows, we elaborate on the process of extracting and subtracting these divergences as appropriate on backgrounds that transition through different epochs. We first focus upon the divergences associated with vacuum two point correlation functions and the associated energy momentum tensors of scalar fields, after which we shift our attention to gravitational waves. In order to do this, we need to establish some preliminary facts and conventions for the discussion that follows.

The energy momentum tensor for a non-interacting, minimally coupled test scalar field $\phi$ on an FRLW background is given by 
\eqn{}{T^\mu_{~\nu} = \partial^\mu\phi\partial_\nu\phi -\frac{1}{2}\delta^\mu_{\nu}\left(g^{\lambda\beta}\partial_\lambda\phi\partial_\beta\phi  + m^2\phi^2\right),}	
from which we can extract 
\eqn{}{-T^0{}_{0} := \rho = \frac{\phi'^2}{2a^2} + \frac{(\nabla\phi)^2}{2a^2} + \frac{m^2}{2}\phi^2,}
where we work in conformal time, and primes denote derivatives with respect to conformal time. The above is a local density that we can rewrite as the coincident limit of a bilinear form as
\eqn{}{\rho := \lim_{y\to x}\rho(\tau;x,y),}
where
\eqn{r2pt}{\rho(\tau;x,y) :=  \frac{1}{2a^2}\left[\phi'(\tau,x)\phi'(\tau,y) + \nabla_x\phi(\tau,x)\cdot\nabla_y\phi(\tau,y)+  m^2a^2\phi(\tau,x)\phi(\tau,y)\right].}

Given the Fourier transform convention for the corresponding field operator $\hat \phi$
\eqn{fourier}{\hat \phi(\tau,x) = \int \frac{d^3k}{(2\pi)^3}\hat\phi(\tau,k)e^{ik\cdot x},}
where the argument distinguishes the field operator in position space from its Fourier component, and the definition\footnote{Below and in the rest of what follows, the equal time expectation values are short hand for in-in correlation functions. For non-interacting test scalar fields evaluated in the adiabatic vacuum, this reduces to Eq. \ref{psdef} and the following.}
\eqn{psdef}{\frac{k^3}{2\pi^2} \langle\hat\phi(\tau,k)\hat\phi(\tau, k')  \rangle := (2\pi)^3\delta^3(k + k')\mathcal P_\phi (\tau, k),} 
we can express the two point correlation function as
\begin{eqnarray}
\nonumber	
	\langle\hat\phi(\tau,x)\hat\phi(\tau,y) \rangle &=& \int \frac{d^3k}{4\pi} \frac{\mathcal P_\phi(\tau, k)}{k^3}e^{ik\cdot (x-y)},\\ \label{2pt0} &=& \int \frac{dk}{k} \mathcal P_\phi(\tau, k)\frac{\sin(kr)}{kr},
\end{eqnarray}
where we have assumed statistical isotropy to perform the angular integrals, and $r := |x-y|$, and will work with the adiabatic vacuum associated with a given background in what follows. For inflationary spacetimes, this will be the usual Bunch Davies vacuum state. The coincident limit of Eq. \ref{2pt0} can be expressed as
\eqn{pscon}{\langle\hat\phi(\tau,x)\hat\phi(\tau,x) \rangle = \int_0^\infty \frac{dk}{k} \mathcal P_\phi(\tau, k),}
and the coincident limit of Eq. \ref{r2pt} can similarly be expressed as  
\eqn{r2pt2}{\lim_{y\to x}\rho(\tau;x,y) = \frac{1}{2 a^2}\int_0^\infty \frac{dk}{k}\left[\left(k^2 + a^2 m^2\right)\mathcal P_\phi(\tau,k) + \mathcal P_{\phi'}(\tau,k)\right],}
where we have used the shorthand
\eqn{}{\frac{k^3}{2\pi^2} \langle\hat\phi'(\tau,k)\hat\phi'(\tau, k')  \rangle := (2\pi)^3\delta^3(k + k')\mathcal P_{\phi'} (\tau, k).}

In terms of canonically normalized mode functions $\phi_k(\tau)$ of the appropriate adiabatic vacuum, the free field operator admits the expansion  
\eqn{}{\hat\phi(\tau,k) = \hat a_{k} \phi_k(\tau) + \hat a^\dag_{-k}\phi_k^*(\tau),}
where $[\hat a_{k} , \hat a^\dag_{-k'}] = (2\pi)^3\delta^3(k + k')$, so that
\eqn{}{\mathcal P_\phi (\tau, k) = \frac{k^3}{2\pi^2}|\phi_k(\tau)|^2,}
and so that Eq. \ref{r2pt2} can be expressed as
\eqn{r2pt3}{\lim_{y\to x}\rho(\tau;x,y) = \frac{1}{4\pi^2 a^2}\int_0^\infty k^2\,dk\left[(k^2 + a^2 m^2)|\phi_k(\tau)|^2 + |\phi'_k (\tau)|^2\right].}
On a purely dS background, the mode functions for a massless field are given by
\eqn{modDS}{|\phi_k(\tau)|^2 = \frac{H^2}{2k^3}(1 + k^2\tau^2),~~~~ |\phi'_k(\tau)|^2 = \frac{H^2}{2k^3}k^4\tau^2,}
so that the late time power spectrum for a massless test scalar is exactly scale invariant with amplitude determined from the Bunch Davies vacuum as $\mathcal P_\phi = \left(\frac{H}{2\pi}\right)^2$, where $H$ is the Hubble factor that defines the dS background, so that Eq. \ref{pscon} reduces to
\eqn{sfd}{\lim_{\tau\to 0}\langle\hat\phi(\tau,x)\hat\phi(\tau,x) \rangle = \left(\frac{H}{2\pi}\right)^2\int_0^\infty \frac{dk}{k}.}
This above is evidently divergent in both the UV and IR. Qualitatively similar divergences are encountered when one constructs the corresponding energy momentum tensor, as we attend to in the next section, warranting a detour on the different approaches one might adopt to regularize expressions like the above where important caveats immediately arise.

\subsection{Regularization scheme (in)dependence}

Given the absence of any scale in the integrand in Eq. \ref{sfd}, the integral vanishes in mass independent regularization schemes such as dimensional or zeta function regularization (see e.g. the relevant chapters of \cite{Kleinert:2001ax, Burgess:2020tbq}). One might perhaps be wary of this and imagine instead regulating such an expression by imposing hard cutoffs in both the UV and the IR. However, one is immediately presented with a choice as to precisely how. For instance,  \cite{Burgess:2009bs} adopts cutoffs in terms of physical momenta on the basis that since IR divergences must cancel in all physical processes, the IR scales associated with these must be expressed in terms of physical scales. On the other hand, \cite{Baumgart:2019clc} argues that it is more practical to adopt cutoffs in terms of comoving momenta for IR divergences given that the existence of a pre-inflationary phase ought to serve as a natural regulator where the distinction becomes less relevant (in the sense that it leads to only sub-leading corrections)\footnote{Logarithmic divergences are special in that they involve the ratios of scales for some observable at some time, where the distinction between comoving and physical cutoffs becomes irrelevant.}. Although we explicitly demonstrate the nature of a pre-inflationary phase as an IR regulator in subsequent sections, we wish to stress that aside from the expected scheme dependence that will drop out of physical quantities once renormalization conditions are imposed consistently\footnote{That is to say, a suitable counterterm can be subtracted that doesn't invalidate the background field quantization prescription one has presumed.}, using hard cutoffs in physical momenta works particularly straightfowardly for matching with the logarithmic UV divergences one can extract from scaleless integrals, and moreover leads to straightforward identification of the requisite counterterms. Simply put, one can arrive at the same expressions for the logarithmic running of physical quantities regardless of regularization scheme with sufficient care, as is to be expected. The fact that UV and IR divergences can be extracted and separated from scaleless integrals may come as a surprise to those used to discarding them, so it behooves us to elaborate on this in detail.  

\subsection{\label{scaleUV} UV and IR divergences in mass (in)dependent schemes}

Reconsider the scaleless integral Eq. \ref{sfd}, which can formally be rewritten as
\eqn{sfd2}{ \int_0^\infty \frac{dk}{k} = \int_0^\infty \frac{k^3 dk}{k^4} = \frac{1}{2\pi^2}\int_{-\infty}^\infty \frac{d^4k}{k^4},}
where the formal manipulations that result in the above render this integral to be Euclidean by default. We see that Eq.\ref{sfd2} is a specific case of the more general form \cite{Burgess:2020tbq}
\eqn{Id}{I_D(m^2) = \int \frac{d^D k}{(2\pi)^D} \frac{k^{2A}}{(k^2 + m^2)^B}.}
Formally, Eq. \ref{Id} evaluates to\footnote{The integral evaluates to $\big|\frac{x^{A + D/2}}{2A + D} {}_2F_1[B,\,A + D/2,\,1 + A + D/2,\,-x]\big|^\infty_0$ with Eq. \ref{Ians} coming from the upper limit along with additional divergent contributions from the upper and lower limits if $A - B + D/2 > 0$ or if $ A +  D/2 < 0$, which are subleading to the divergences in Eq. \ref{Ians} for integer values of $A$ and $B$ as $D\to 4$.}
\eqn{Ians}{I_D(m^2) = \frac{\Gamma\left(A+\frac{D}{2}\right) \Gamma\left(B-A-\frac{D}{2}\right)}{(4\pi)^{D/2}\Gamma\left(\frac{D}{2}\right)\Gamma(B)}\left(m^2\right)^{A-B+D/2},}
which permits analytic continuation to non-integer values of $D$. Re-expressing the two point function Eq. \ref{sfd} via Eq. \ref{sfd2} as\footnote{Dimensional regularization requires us to deform the dimension of every quantity that depends on spacetime dimension, including the mode functions on the background in question. This is especially crucial in the context of loop corrections in the presence of interactions \cite{Senatore:2009cf, delRio:2018vrj, Baumgart:2021ptt}, where spuriously large loop corrections might otherwise be inferred. However, because we only consider bubble diagrams of non-interacting fields in what follows (with no external momenta), the result of doing so contributes only additional finite contributions and will be elided in what follows.}
\eqn{sfd3}{\lim_{\tau\to 0}\langle\hat\phi(\tau,x)\hat\phi(\tau,x) \rangle = \frac{H^2}{8\pi^4} \int_{-\infty}^\infty \frac{d^4k}{k^4} = \frac{H^2}{8\pi^4} \int_{-\infty}^\infty d^4k\left[\frac{1}{k^2(k^2 + m^2)} + \frac{m^2}{k^4(k^2 + m^2)} \right],}
we see via Eq. \ref{Ians} that this evaluates in $D = 4 - \delta$ dimensions to two equal and opposite contributions of the form
\eqn{}{\pm\frac{H^2}{4\pi^2}\left[\frac{1}{\delta} - \frac{1}{2}\left(\log \frac{m^2}{4\pi\mu^2} + \gamma_E - 1 \right) \right],}
where $\gamma_E$ is the Euler-Mascheroni constant, and $\mu$ is some arbitrary mass scale necessitated by dimensional deformation. The sum of the two contributions can be written as
\eqn{drl}{\lim_{\tau\to 0}\langle\hat\phi(\tau,x)\hat\phi(\tau,x) \rangle = \left(\frac{H}{2\pi}\right)^2 \left[\frac{1}{\delta_{\rm UV}} - \frac{1}{\delta_{\rm IR}}  + \log \frac{\mu_{\rm UV}}{\mu_{\rm IR}}\right],}
where we have artificially given a separate label to the dimensional deformation parameters $\delta$ and $\mu$ from the two terms to distinguish the UV and IR divergent contributions in Eq. \ref{sfd3}, illustrating the anatomy of how dimensional regularization works in canceling two separately divergent contributions by default. 

It is informative to compare this result with the outcome of imposing hard cutoffs. Whether we chose to do so in Eq. \ref{sfd} in terms of physical momenta (where $\Lambda_{\rm UV/ IR} = a\, k_{\rm UV/ IR}$) or in terms of comoving momenta (where $\Lambda_{\rm UV/ IR} = k_{\rm UV/ IR}$) is immaterial in the context of logarithmic divergences which only sees ratios of these scales:  
\eqn{cd}{\lim_{\tau\to 0}\langle\hat\phi(\tau,x)\hat\phi(\tau,x) \rangle = \left(\frac{H}{2\pi}\right)^2\log \left(\frac{k_{\rm UV}}{k_{\rm {IR}}}\right).}
From this, we see that in spite of the fact that scaleless integrals vanish under dimensional regularization, factorizing them into a sum of scaleful integrals allows one to match the coefficients of the UV divergent logarithms in Eq. \ref{drl} with those in Eq. \ref{cd} obtained by imposing hard cutoffs in physical momenta. As we show in subsequent sections, this will also be true no matter the background. 

\section{Non-interacting test scalar fields}

In this section, we examine divergences in the two point correlation function and energy momentum tensor for non-interacting, minimally coupled massless test scalar fields on dS and quasi dS backgrounds, as well as that for sufficiently light massive scalar fields on a dS background. All divergent integrals turn out to be scaleless, a feature that persists even when one considers a background that transitions from a pre-inflationary epoch through to inflation and exiting to a terminal phase of radiation domination. Nevertheless, one can extract UV divergences via the techniques elaborated in section \ref{scaleUV} and compare with what one would have obtained with a hard cutoff in physical momenta, recovering identical coefficients for the logarithmic divergences. 

\subsection{\label{sec:qdssf} Divergences -- (quasi) dS backgrounds}

On backgrounds that deviate from dS, one has to be careful to factor in the difference in the scale factor from the pure dS form of $a(\tau) = -1/(H\tau)$. For constant but non-zero $\epsilon := -\dot H/H^2$, as is the case during power law inflation for instance, one has $a(\tau) = 1/[-\tau H_0]^{\nu-1/2}$, where $\nu = \frac{3 -\epsilon}{2(1-\epsilon)}$ (and where the integration constant $H_0$ corresponds to the value of $H$ at $t = 0$ in cosmological time -- cf. \cite{Chluba:2015bqa}). The net result for the coincident limit of the two point function for a massless scalar field on a background corresponding to constant $\epsilon$ is given by
\eqn{}{\lim_{\tau\to 0}\langle\hat\phi(\tau,x)\hat\phi(\tau,x) \rangle = \left(\frac{H_0}{2\pi}\right)^2 \frac{\Gamma^2(\nu)}{\pi}2^{2\nu-1}  \int_0^\infty \frac{dk}{k} \left(\frac{k}{H_0}\right)^{3 -2\nu}\left[1 + \frac{k^2\tau^2}{2(\nu-1)} + ... \right],}
or more generally,
\eqn{qdsdiv}{\lim_{\tau\to 0}\langle\hat\phi(\tau,x)\hat\phi(\tau,x) \rangle \sim \left(\frac{H_0}{2\pi}\right)^2\int_0^\infty \frac{dk}{k} \left(\frac{k}{H_0}\right)^{n_s-1}\left[1 + \left(\frac{k}{a H_0}\right)^2\frac{a^{n_s-1}}{2 - n_s} + ...\right]}
where the ellipses denote terms that vanish in the $\tau \to 0$ limit and the $\sim$ indicates a numerical pre-factor that is close to unity for $n_s \approx 1$ (and tends to it in the dS limit). In the dS limit, we find the relatively simple expression
\eqn{dsdiv0}{\lim_{\tau\to 0}\langle\hat\phi(\tau,x)\hat\phi(\tau,x) \rangle = \left(\frac{H_0}{2\pi}\right)^2\int_0^\infty \frac{dk}{k} \left[1 + \left(\frac{k}{a H_0}\right)^2 \right],}
which supplements Eq. \ref{sfd} with a sub-leading correction in the late time limit, notable relative to the analogous expression on quasi dS Eq. \ref{qdsdiv} in that Eq. \ref{dsdiv0} is exact for all times. The scaleless nature of Eq. \ref{qdsdiv} implies that it vanishes when regularized in any mass independent scheme. Instead, one might contemplate imposing a hard UV or IR cutoff in physical momenta by setting $k_{\rm UV/IR} = a \Lambda_{\rm UV/IR}$, and subtracting the UV divergence with the necessary counterterm. Doing so for the expression Eq. \ref{dsdiv0}, one finds (see e.g. \cite{Burgess:2009bs})
\begin{eqnarray}
\nonumber \lim_{\tau\to 0}\langle\hat\phi(\tau,x)\hat\phi(\tau,x) \rangle &=& \left(\frac{H_0}{2\pi}\right)^2\left[ \int_{a \Lambda_{\rm IR}}^{a \Lambda_{\rm UV}} \frac{dk}{k} \left(1 + \left(\frac{k}{a H_0}\right)^2 \right) \right. \\ \nonumber
&& \left. - \int_{a \mu}^{a \Lambda_{\rm UV}} \frac{dk}{k} \left(1 + \left(\frac{k}{a H_0}\right)^2 \right) \right]\\ 
\label{dsdiv00}  &=& \left(\frac{H_0}{2\pi}\right)^2\left\{\log\left(\frac{\mu}{\Lambda_{\rm IR}}\right) + \frac{1}{2H_0^2}\left(\mu^2 - \Lambda_{\rm IR}^2\right) \right\},~~(n_s = 1)
\end{eqnarray}
where $\mu$ is some arbitrary renormalization scale.  We note that the counterterm that subtracts the UV divergence in Eq. \ref{dsdiv00} is given by 
\eqn{dsct}{{\rm c.t.} = \left(\frac{H_0}{2\pi}\right)^2\left[ \log\left(\frac{\mu}{\Lambda_{\rm UV}}\right) + \frac{1}{2H_0^2}\left(\mu^2 - \Lambda_{\rm UV}^2 \right) \right],}
corresponding to a scheme dependent renormalization of the cosmological constant\footnote{We note that imposing a comoving cutoff in the limits of Eq. \ref{dsdiv00} instead of a physical cutoff would have resulted in a time dependence in the required counterterms (at odds with the dS invariance of the vacuum state) and would have necessitated further gymnastics to arrive at any physically meaningful quantity.}. That these expressions are independent of time is a consequence of having imposed a physical cutoff on a dS invariant background. The remaining IR divergence is indicative of the fact that we have not yet arrived at something that can be processed into an observable quantity. It could end up being canceled by other contributions when we compute a well defined observable, or it could herald the need for resummation in which case it will also eventually get canceled. It could also indicate that the background we are attempting quantization around is not what it seems \cite{Creminelli:2008es}, and additional physical inputs are required in order to proceed. As we will show explicitly in what follows, were we to abandon the assumption that inflation was past dS eternal, then all IR divergences cancel among themselves and are in effect regulated by the scale corresponding to the beginning of inflation.

It is informative to compare what we would have obtained if we had dimensionally regularized Eq. \ref{dsdiv0} instead. We first note that the scaleless nature of the integral permits us to change variables to $q:= k/(a H_0)$ for any fixed time. Hence the integral we need to evaluate can be factorized as in Eq. \ref{sfd3} as:
\eqn{dsdr0}{\frac{H^2_0}{8\pi^4}\int_{-\infty}^\infty \frac{d^4 q}{q^4}(1 + q^2) = \frac{H_0^2}{8\pi^4} \int_{-\infty}^\infty d^4q\left[\frac{1}{(q^2 + \widetilde m^2)} + \frac{1 + \widetilde m^2}{q^2(q^2 + \widetilde m^2)} + \frac{\widetilde m^2}{q^4(q^2 + \widetilde m^2)} \right]}
where $\widetilde m := \mu/H_0$ is some auxiliary dimensionless mass scale. The first and second terms above are power law and log divergent in the UV, respectively, whereas the third term is IR divergent. The UV divergences must of course, be subtracted by an appropriate counterterm. One finds after cancellations between the contributions from the first and second terms of Eq. \ref{dsdr0}, a remaining UV divergence 
\eqn{}{{\rm c.t.} = \left(\frac{H_0}{2\pi}\right)^2\left\{\log\left(\frac{ \mu}{H_0}\right) - \frac{1}{\delta_{\rm UV}} + \frac{1}{2}\left(\gamma_E  - 1 - \log 4\pi \right)\right\},}
which is to be compared to Eq. \ref{dsct}.

The situation for quasi dS is more complicated if one were to regularize it in a mass dependent scheme, as is evident from Eq. \ref{qdsdiv}. Nevertheless, it is still worth elaborating upon provided we simplify matters by further specifying that the background corresponds to that of eternal power-law inflation, defined by a constant but non-zero $\epsilon$, so that $n_s - 1 = -2\epsilon - 2\epsilon^2 + ...$ . We will address the more realistic case of finite duration inflation next. One finds from inspection of the integrand of Eq. \ref{qdsdiv}:
\eqn{}{ \int_0^\infty \frac{dk}{k} \left(\frac{k}{H_0}\right)^{n_s-1}\left[1 + \left(\frac{k}{a H_0}\right)^2\frac{a^{n_s-1}}{2 - n_s}\right]}
that the first term results
\eqn{}{\int_0^\infty \frac{dk}{k} \left(\frac{k}{H_0}\right)^{n_s-1} = \frac{1}{| n_{s}-1 |}\left(\frac{a\Lambda}{H_0}\right)^{n_{s }-1},}
and the second term results
\eqn{}{\int_0^\infty \frac{dk}{k} \left(\frac{k}{a H_0}\right)^{n_s+1}\frac{a^{2(n_s-1)}}{2 - n_s} = \frac{a^{2(n_s-1)}}{(2- n_s)(n_{s }+1)}\left(\frac{\Lambda}{H_0}\right)^{n_{s }+1}}
where whether $\Lambda$ is an IR or UV cutoff depends on the value of $n_s$. By rewriting Eq. \ref{qdsdiv} as a function of the slow roll parameter $\epsilon$ 
\eqn{qdsdrint}{\lim_{\tau\to 0}\langle\hat\phi(\tau,x)\hat\phi(\tau,x) \rangle \sim \left(\frac{H_0}{2\pi}\right)^2 \int_0^\infty \frac{dk}{k} \left(\frac{k}{H_0}\right)^{-2 \epsilon}\left[1 + \left(\frac{k}{a H_0}\right)^2\frac{a^{-2 \epsilon}}{1+2 \epsilon}\right]}
we find that in the quasi dS limit ($\epsilon \ll 1$) the first term gives a logarithmic divergent contribution
\eqn{nsI}{ \int_0^\infty \frac{dk}{k} \left(\frac{k}{H_0}\right)^{-2 \epsilon} = \frac{1}{ 2 \epsilon } +\log\frac{\Lambda_{\rm UV}}{\Lambda_{\rm IR}} +\mathcal{O}(\epsilon)}
and the second term a power law UV divergence
\eqn{}{\int_0^\infty \frac{dk}{k} \left(\frac{k}{a H_0}\right)^{2(1-\epsilon)}\frac{a^{-4 \epsilon}}{1+2 \epsilon} = \frac{a^{-4 \epsilon}}{2 }\left(\frac{\Lambda_{\rm UV}}{H_0}\right)^{2}.}
The interpretation of the IR divergences is as discussed above while the UV divergence must be subtracted with a local counterterm if we are to obtain physically meaningful quantities. For small $\epsilon$, the counterterm is given by
\begin{eqnarray} \nonumber
{\rm c.t.} &&= \lim_{\tau\to 0} \left(\frac{H_0}{2\pi}\right)^2 \left[ \frac{1}{2 \epsilon} + \log\frac{\mu}{\Lambda_{\rm UV}} + \frac{a^{-4 \epsilon}}{2 }\left(\left(\frac{\mu}{H_0}\right)^{2}  -  \left(\frac{\Lambda_{\rm UV}}{H_0}\right)^{2}\right) + \mathcal{O}(\epsilon) \right] \\ \label{qdsfinite}
&& = \lim_{\tau\to 0} \left(\frac{H_0}{2\pi}\right)^2 \left[ \frac{1}{2 \epsilon} + \log\frac{\mu}{\Lambda_{\rm UV}} - \frac{a^{-4\epsilon} }{2} \frac{\Lambda^2_{\rm UV}}{H_0^2} + \mathcal{O}(\epsilon) \right]
\end{eqnarray}
where again $\mu$ is some renormalization scale and given that $\Lambda_{\rm UV} \gg \mu$, we have neglected the sub-leading dependence on $\mu$ in the second equality. We first note that the required counterterm is time dependent, which should not come as a surprise given that the background is no longer maximally symmetric. It behooves us to elaborate on the specific form of the local counterterm that could possibly have this particular time dependence. Noting that on the specified power law inflating background, $\dot H, H^2 \sim a^{-2 \epsilon}$, so that curvature squared invariants evaluated on this background have the required secular dependence. That is, Eq. \ref{qdsfinite} derives from
\eqn{}{{\rm c.t.} \subset \int d^4 x\sqrt{-g}\left[c_1(\mu)R^2 + c_2(\mu)R_{\mu\nu}R^{\mu\nu} \right],}
which are the usual leading curvature squared counterterms encountered in the effective field theory treatment of gravity \cite{Donoghue:1994dn, Donoghue:1995cz, Donoghue:2012zc, Burgess:2003jk}.

As before, we can retrace the same computation using dimensional regularization. We first consider the separate contributions to the integrand Eq. \ref{qdsdrint}. The logarithmic divergent contribution can be re-expressed as
\begin{eqnarray} \nonumber
\int_0^\infty \frac{dk}{k} \left(\frac{k}{H_0}\right)^{-2\epsilon} &=& \frac{1}{2\pi^2}\int_{-\infty}^\infty \frac{d^4 q}{q^4} q^{-2\epsilon} \\ \label{IRdr}
&=& \frac{1}{2\pi^2} \int_{-\infty}^\infty d^4q\left[\frac{q^{-2\epsilon}}{q^2(q^2 + \widetilde m^2)} + \frac{\widetilde m^2 q^{-2\epsilon}}{q^4(q^2 + \widetilde m^2)} \right],
\end{eqnarray}
where we have switched to the dimensionless variable $q := k/H_0$, and similarly for $\widetilde m := \mu/H_0$. By isolating the UV pole, we find that the presence of the non-integer exponents eliminates the $\delta$ poles and we obtain that in the quasi dS limit the first term results \footnote{Note that we are required to take the limits in the order $\delta \to 0$ and then $\epsilon \to 0$. To do the opposite would mean to have regularized on a dS background first and then deformed the background in the hopes that no new divergences will have appeared through this process. One can show that this is not justified on general grounds \cite{Birrell:1982ix}, and is seen directly from the proliferation of higher order $\delta$ poles encountered when taking the limits in the opposite order.}
\eqn{}{\left(\frac{H_0}{2\pi}\right)^2 \int_0^\infty \frac{dk}{k} \left(\frac{k}{H_0}\right)^{-2 \epsilon}=\left(\frac{H_0}{2\pi}\right)^2 \left[\frac{1}{2 \epsilon} - \log\frac{\mu}{H_0} + \frac{\epsilon}{12} \left(\pi^2 +24\log \frac{\mu}{H_0} \right)  \right].}
The second term of Eq. \ref{qdsdrint} can be rewritten as
\eqn{UVdr}{\begin{aligned} 
\frac{a^{-4 \epsilon}}{1+2 \epsilon} \int_0^\infty \frac{dk}{k} \left(\frac{k}{a H_0}\right)^{2(1-\epsilon)} &= \frac{a^{-4 \epsilon}}{1+2 \epsilon} \frac{1}{2\pi^2}\int_{-\infty}^\infty \frac{d^4 q}{q^4} q^{2(1-\epsilon)} \\ 
&= \frac{a^{-4 \epsilon}}{1+2 \epsilon} \frac{1}{2\pi^2} \int_{-\infty}^\infty d^4q\left[\frac{q^{2(1-\epsilon)}}{q^2(q^2 + \widetilde m^2)} + \frac{\widetilde m^2 q^{2(1-\epsilon)}}{q^4(q^2 + \widetilde m^2)} \right],
\end{aligned}}
where just as we did on dS space, we work with dimensionless variables $q := k/aH_0$ and $\widetilde m := \mu/H_0$. Using Eq. \ref{Ians} we see that power law UV divergences cancel among themselves, and so do not necessitate any counterterms. In conclusion, we find that the counterterm in the quasi dS limit results 
\begin{eqnarray} 
{\rm c.t.} && = \lim_{\tau\to 0} \left(\frac{H_0}{2\pi}\right)^2 \left[  \frac{1}{2 \epsilon} + \log\frac{\mu}{H_0}  + \mathcal{O}(\epsilon) \right].
\end{eqnarray}

The generalization to massive fields is straightforward on dS backgrounds, where analytic expressions for the mode functions can be obtained: 
\eqn{mfmf}{|\phi_k(\tau)|^2 = \frac{\pi}{4} H_0^2 (-\tau)^3 |H^{(1)}_{\nu_m}(-k \tau)|^2,} 
where 
\eqn{nusq}{\nu^2_m := \frac{9}{4} - \frac{m^2}{H_0^2},}
and where $H^{(1)}_{\nu_m}$ is the corresponding Hankel function with degree $\nu_m$. Presuming $0 < m^2 \leq 9H_0^2/4$ so that $\nu_m$ is still real, the coincident limit of the two point function at late times can then be computed as
\eqn{sfdms}{\lim_{\tau\to 0}\langle\hat\phi(\tau,x)\hat\phi(\tau,x) \rangle = \frac{H_0^2 2^{2\nu_m}\Gamma^2(\nu_m) }{8\pi^3}\int_0^\infty \frac{dk}{k}\left[\left(\frac{k}{a H_0}\right)^{3 - 2\nu_m} + \frac{\left(k/a H_0 \right)^{5 - 2\nu_m}}{2\left(\nu_m-1\right)}\right].}

It is notable that in spite of the presence of the additional mass scale $m$, the integral remains scaleless\footnote{This can be understood from the fact that on dS, the quantity nominally labeled mass in Eq. \ref{nusq} is constructed from eigenvalues of the Casimir operators corresponding to conformal dimension and spin expressed in units of the dS radius, which remains the only scale in the problem. We thank Paolo Benincasa for pointing this out to us.}. Here also, we can make the change of variable to $q = k/(aH)$, so that the above can be recast as
\begin{eqnarray}
\label{sfdms}\lim_{\tau\to 0}\langle\hat\phi(\tau,x)\hat\phi(\tau,x) \rangle &=& \frac{H_0^2}{16\pi^5}2^{2\nu_m}\Gamma^2(\nu_m) \int_{-\infty}^\infty \frac{d^4q}{q^4} q^{3 - 2\nu_m}\left[1 + \frac{q^{2}}{2(\nu_m - 1)}\right], 
\end{eqnarray}
where the first thing to note is that the IR divergence encountered for a non-interacting massless scalar on dS is eliminated given that $\nu_m < 3/2$\footnote{A short-lived conclusion on dS backgrounds the minute any interactions are incorporated.}. Proceeding exactly as before, we can compare the UV divergence obtained from imposing a hard cutoff in physical momenta to that obtained by factorizing the scaleless integral in dimensional regularization. In the limit $m^2/H_0^2 \ll 1$, one finds that the required counterterms are given by
\eqn{}{{\rm c.t.} = \frac{H_0^2}{8\pi^2}\frac{2^{2\nu_m}\Gamma^2(\nu_m)}{\pi}\left[\frac{3H_0^2}{m^2} + \log\frac{\mu}{\Lambda_{\rm UV}} -\frac{\Lambda_{\rm UV}^2}{4 H_0^2} + ... \right]~~~~~~~{(\rm physical ~cutoff)}}
and
\eqn{}{{\rm c.t.} = \frac{H_0^2}{8\pi^2}\frac{2^{2\nu_m}\Gamma^2(\nu_m)}{\pi}\left[\frac{3H_0^2}{m^2} + \log\frac{\mu}{H_0} + ... \right]~~~~~~~~~~~{(\rm dim~ reg)} }
where again, one takes the limit $\delta \to 0$ to regularize the integrals before taking the limit $m^2/H_0^2 \to 0$. In both cases, the required counterterm corresponds to a renormalization of the cosmological constant.  

Having labored over the details of regularizing and renormalizing divergences for the two point function of test scalar fields, repeating the exercise for the associated stress tensor may seem a straightforward extension. However, we immediately encounter a difficulty: it is not always possible to construct a counterterm that subtracts the UV divergences encountered from background geometric invariants if the regularization scheme itself does not respect this symmetry. This is not to a concern to be casually dismissed\footnote{An issue that did not arise for the two-point function given that its nature as a scalar bilinear.}, as persisting with this regularization scheme could lead us to the erroneous conclusion that the background we attempted to quantize around is transmuting into something else under renormalization if not worked through to the end. It is here that dimensional and related zeta function regularization techniques distinguish themselves in theories with general covariance, as they effortlessly preserve the symmetries of the background even if the precise mechanism by which this occurs can seem non-trivial. For this reason, it is worth explicitly tracing through this process in detail\footnote{Nevertheless, it remains true that the coefficients of the UV divergent logarithms would match regardless of whether one dimensionally regularizes or one imposes hard cutoffs in physical momenta.}. 

Consider a minimally coupled non-interacting test scalar field on dS space in the Bunch Davies vacuum. The rotational invariance of the vacuum state allow us to write the non-vanishing components as
\eqn{rmfds}{\rho  = \frac{1}{4\pi^2 a^2}\int_0^\infty k^2\,dk\left[\left(k^2 + a^2 m^2\right)|\phi_k(\tau)|^2 + |\phi'_k (\tau)|^2\right],}
and
\eqn{pmfds}{p  = \frac{1}{4\pi^2 a^2}\int_0^\infty k^2\,dk\left[-\left(\frac{k^2}{3} + a^2 m^2\right)|\phi_k(\tau)|^2 + |\phi'_k (\tau)|^2\right],}
with all other components of the energy momentum tensor vanishing. From the asymptotic forms of Eq. \ref{mfmf}, one can infer that both integrals are UV divergent. The energy density in the massless limit is straightforwardly computed via Eq. \ref{modDS}, and found to be
\eqn{rdshc}{\rho = \frac{H^4}{8\pi^2}\int^\infty_0 \frac{dk}{k} \left[ \left(\frac{k}{a H}\right)^2 + 2\left(\frac{k}{a H}\right)^4\right],}
and the corresponding pressure is found to be 
\eqn{}{p = \frac{H^4}{8\pi^2}\int^\infty_0 \frac{dk}{k} \left[ -\frac{1}{3}\left(\frac{k}{a H}\right)^2 + 2\left(\frac{k}{a H}\right)^4\right],}
Although one can regularize both components through imposing physical cutoffs, one is immediately confronted by the fact that the required counterterm for the leading divergences will not be proportional to the metric nor the Ricci tensor on the presumed dS background (which would correspond either to renormalizations of the cosmological and Newton's constants). This is not the case when one implements dimensional regularization, which may seem rather non-trivial considering the forms that Eqs. \ref{rmfds} and \ref{pmfds} take. Although a fully covariant formalism to regularize stress tensors using mass independent schemes is the industry standard (as elaborated upon in \cite{Birrell:1982ix, Parker:2009uva}), it is nevertheless informative to trace through this process for the foliation specific forms Eq. \ref{rmfds} and \ref{pmfds}, as we do in Appendix \ref{app:emt}.

The important lesson to draw from this is that regularizing the divergence of just one particular component of the stress tensor (such as the energy density) is not enough. Computing physically sensible quantities derived from the energy momentum tensor in a reliable manner necessitates constructing the appropriate counterterm, without which renormalization conditions cannot be consistently imposed. If the required counterterm cannot be constructed from background quantities in a particular regularization scheme, then all subsequent conclusions will necessarily be scheme dependent, and therefore unphysical. For this reason, regularization schemes that preserve general covariance such as dimensional or zeta function regularization are preferred in this context. In spite of this caveat, however, one can still be forgiven for using hard cutoffs in physical momenta when restricted to calculating logarithmic divergences, as one can show the result to be identical to what would have been obtained in a mass independent scheme. We will return to this point in regards the regularization of the stress tensor for gravitational waves in the next section.   

\subsection{\label{sec:fd} Divergences -- scalar fields and finite duration inflation}
Having acquainted ourselves with the basics of regularizing and renormalizing divergences for familiar quantities, one would like to generalize this to backgrounds that realistically model the universe we inhabit. In particular, one might wonder if the fact that we only obtained scaleless integrals in the previous subsections is an artifact of considering the unrealistic scenario of infinite duration inflation. We show that this situation persists even when we consider a cosmology that transitions between different epochs. 

In order for inflation to have finite duration by definition, we must be considering an epoch before and after inflation, which we choose to be radiation domination in both cases. We consider the scale factor to evolve as

\begin{eqnarray}
\nonumber 
	a(\tau) &=& a_{\rm R} \left(2 - \frac{\tau}{\tau_{\rm I}}\right)e^{-\mathcal N_{\rm tot}}~~~~ \tau < \tau_{\rm I}\\ \nonumber &=& a_{\rm R}\left(\frac{\tau_{\rm I}}{\tau}\right)e^{-\mathcal N_{\rm tot}}~~~~ \tau_{\rm I} < \tau < \tau_{\rm R}\\ \label{sfdef} &=& a_{\rm R}\left(2 - \frac{\tau}{\tau_{\rm R}} \right) ~~~~~~ \tau_{\rm R} < \tau
\end{eqnarray}
where $a_{\rm R}$ is the scale factor at reheating, and $\tau_{\rm I}$ and $\tau_{\rm R}$ correspond to the (negative) conformal time at the start and end of inflation\footnote{Note that this puts the initial singularity at $\tau = 2\tau_{\rm I}$, however the only manner in which the pre-inflationary phase bears on late time observables is through the choice to begin with the adiabatic vacuum evolved to the start of the inflationary epoch at $\tau_{\rm I}$.} respectively, and 
\eqn{}{\mathcal N_{\rm tot} = \log (a_{\rm R}/a_{\rm I}) = \log (\tau_{\rm I}/\tau_{\rm R})}
is the total amount e-folds of inflation. The domain of $\tau$ is $(-\infty,\infty)$ with inflation occurring during negative conformal time. With these definitions, the Hubble rate during inflation is given by
\eqn{Hinfdef}{H = -\frac{1}{a_{\rm R}\tau_{\rm R}}.}
In what follows, we keep the normalization $a_{\rm R}$ arbitrary, although one can readily set $a_{\rm R} \equiv 1$ for convenience in what follows. Here again, we focus on a massless, minimally coupled non-interacting scalar field for illustrative purposes. The mode functions during the terminal stage of radiation domination can be rewritten for the purposes of a matching to the end of inflation as
\eqn{modRD2}{\begin{aligned}
\phi^{\rm RD}_k &= \frac{1}{a}\frac{1}{\sqrt{2k}}\left[\alpha_k^{\rm R} e^{-i k \tau_{\rm R} \left( 2- \frac{a}{a_{\rm R}}\right)} + \beta_k^{\rm R} e^{i k \tau_{\rm R} \left( 2- \frac{a}{a_{\rm R}}\right)}  \right] \\ 
	\phi'^{\rm RD}_k &= \frac{a_{\rm R}}{a^2\tau_{\rm R}}\frac{1}{\sqrt{2k}}\left[\alpha_k^{\rm R} e^{-i k \tau_{\rm R} \left( 2- \frac{a}{a_{\rm R}}\right)}\left(1 - i k \tau_{\rm R} \frac{a}{a_{\rm R}}\right) + \beta_k^{\rm R} e^{i k \tau_{\rm R} \left( 2- \frac{a}{a_{\rm R}}\right)}\left(1 +i k \tau_{\rm R} \frac{a}{a_{\rm R}}\right)  \right].
\end{aligned}}
The mode functions during inflation are given by
\begin{eqnarray}
\nonumber
	\phi_k^{\rm I} &=& \frac{H}{\sqrt{2k^3}}\left[\alpha_k^{\rm I} e^{i \frac{k}{aH}}\left(1 - \frac{i k}{a H}\right) +  \beta_k^{\rm I} e^{-i \frac{k}{aH}}\left(1 + \frac{i k}{a H}\right)\right] \\ 
	\phi'^{\rm I}_k &=& -\frac{H}{\sqrt{2k^3}}\left[\alpha_k^{\rm I} e^{i \frac{k}{aH}}\frac{k^2}{a H} +  \beta_k^{\rm I} e^{-i \frac{k}{aH}}\frac{k^2}{aH}\right],
\end{eqnarray}
where the presence of non-trivial Bogoliubov coefficients come from having matched to a radiation dominated pre-inflationary phase initiated in the adiabatic vacuum, where the mode functions are given by 

\begin{eqnarray}
\nonumber
	\phi^{\rm PI}_k &=& \frac{1}{a}\frac{1}{\sqrt{2k}}e^{-i k \tau_{\rm I} \left( 2 - \frac{a}{a_{\rm R}} e^{\mathcal N_{\rm tot}}\right)} \\ 
	\phi'^{\rm PI}_k &=& -\frac{1}{a^2\sqrt{2k}} \frac{a_{\rm I}}{\tau_{\rm I}} e^{-i k \tau_{\rm I} \left( 2 - \frac{a}{a_{\rm R}} e^{\mathcal N_{\rm tot}}\right) }\left(-1 + i k \tau_{\rm I} \frac{a}{a_{\rm I}} \right).
\end{eqnarray}
One first obtains $\alpha^{\rm I}_k$ and $\beta_k^{\rm I}$ by matching to the pre-inflationary radiation dominated epoch, so that
\begin{eqnarray} \nonumber
	\alpha_k^{\rm I} &=& \left(i- i \frac{a_{\rm I}^2  H^2}{2k^2}+ \frac{a_{\rm I}  H}{k}  \right) \\ \label{mcI2}
	\noindent  \beta_k^{\rm I} &=& i \frac{a_{\rm I}^2 H^2}{2k^2}  e^{2i\frac{k}{a_{\rm I} H} }.
\end{eqnarray}
Similarly, after the matching between inflation and radiation dominated era, one finds that $\alpha^{\rm R}_k$ and $\beta_k^{\rm R}$ are given by
\begin{eqnarray}
	\label{mcR}
	\nonumber  \alpha_k^{\rm R} &=&  \alpha_k^{\rm I} \left(-i+ \frac{a_{\rm R}   H}{k}+i \frac{a_{\rm R} ^2  H^2}{2k^2}  \right)  +i \beta_k^{\rm I}   \frac{a_{\rm R} ^2  H^2}{2k^2}  e^{-2i\frac{k}{a_{\rm R}  H}}  \\
	\beta_k^{\rm R}  &=&-i \alpha_k^{\rm I} \frac{a_{\rm R} ^2  H^2}{2k^2}  e^{2i\frac{k}{a_{\rm R}  H}} + \beta_k^{\rm I}  \left(i+ \frac{a_{\rm R}   H}{k}-i \frac{a_{\rm R} ^2  H^2}{2k^2}  \right).
\end{eqnarray}

\begin{figure}
 \begin{subfigure}{.54\textwidth}
	\begin{flushleft}
	 \hspace{-1.3cm}
	\includegraphics[width=3.2in]{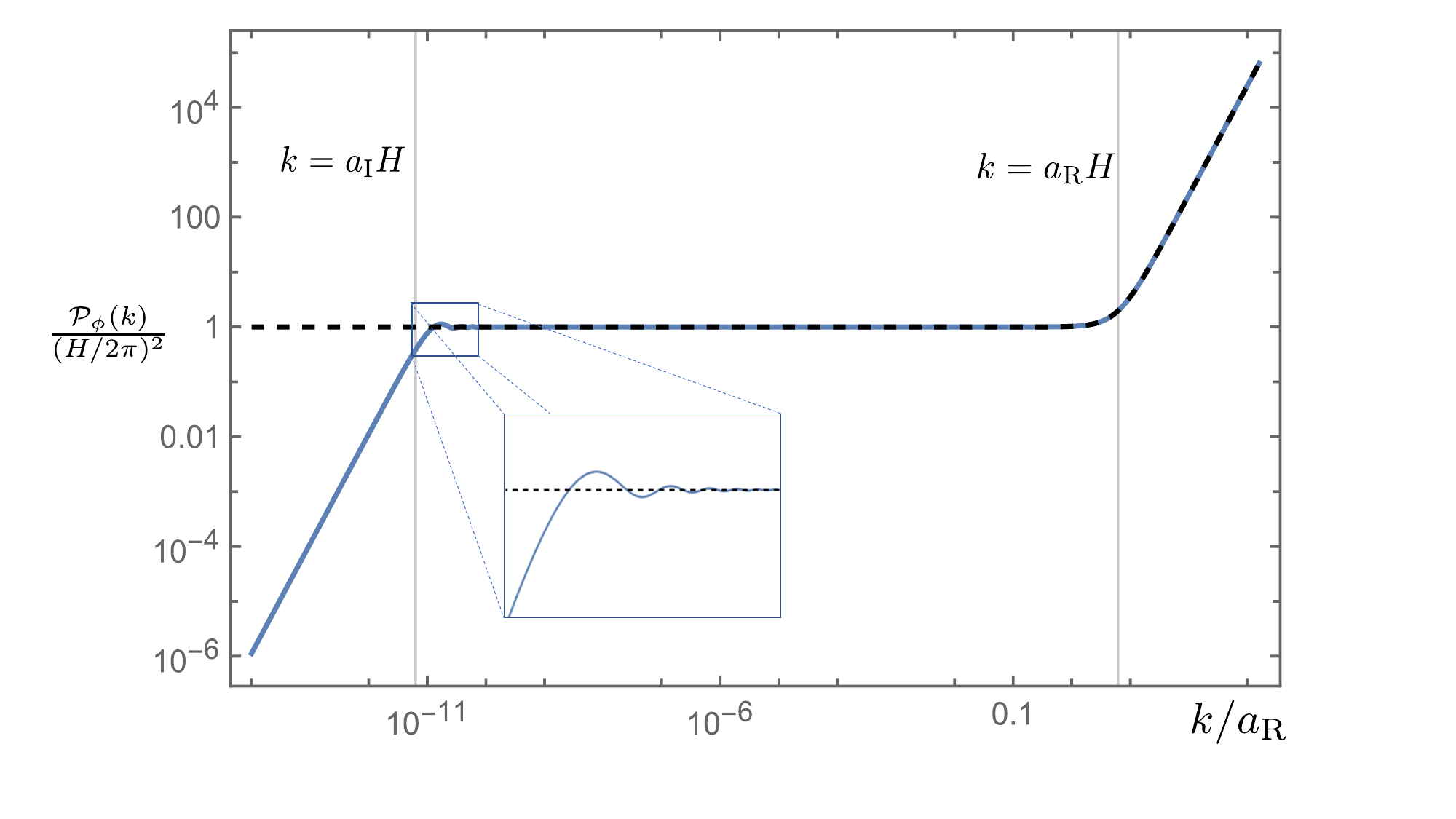}  
	\caption{Power spectra evaluated at reheating $a = a_{\rm R}$, where $a_{\rm I} = 10^{-12} a_{\rm R}$ in units where $H$  is set to $2\pi$.}
	\label{SUBFIGURE LABEL 1}
	\end{flushleft}
 \end{subfigure}
 \begin{subfigure}{.54\textwidth}
  \hspace{-1.3cm}
	\centering
	\includegraphics[width=3.2in]{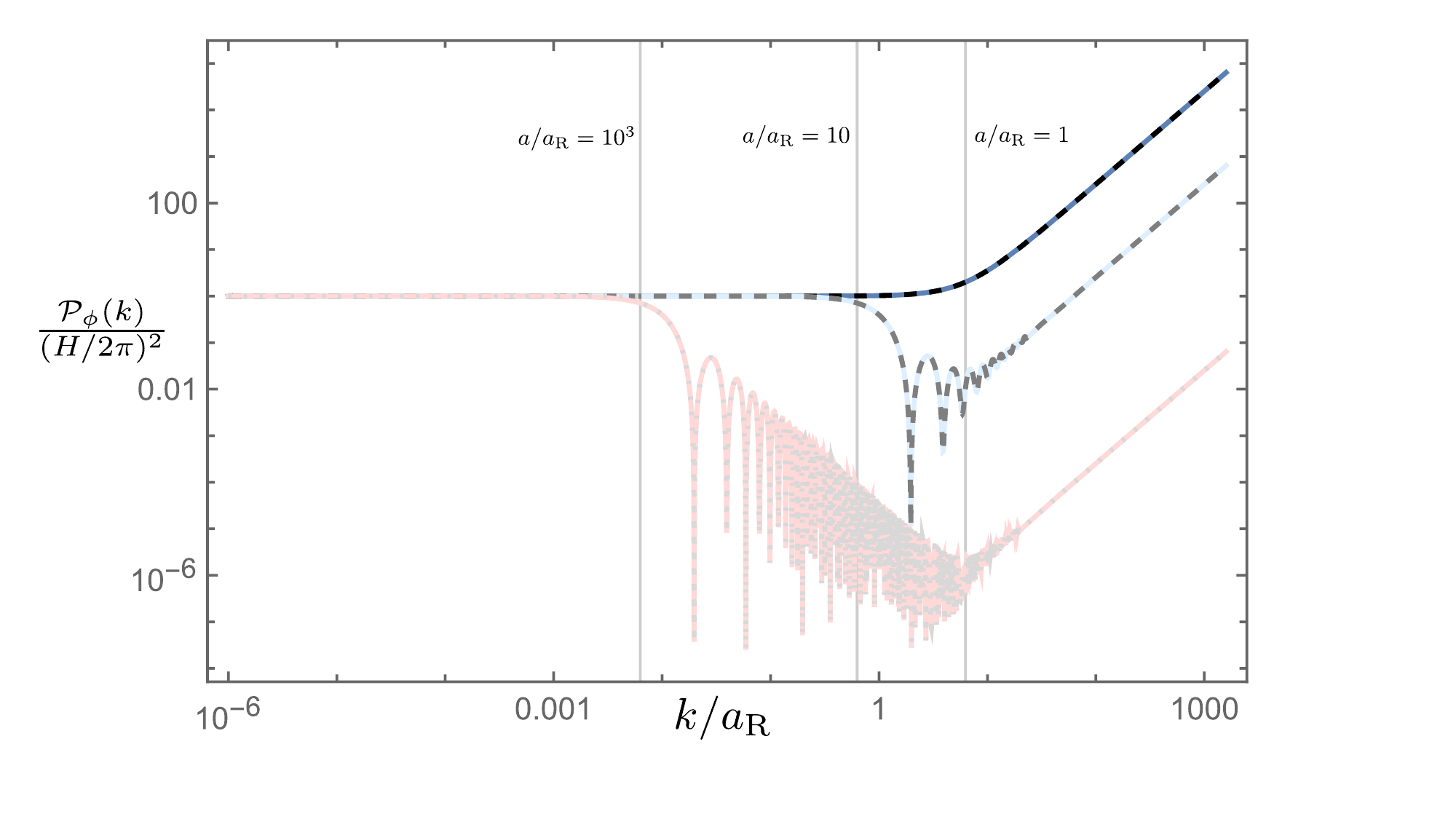}  
	\caption{Power spectra evaluated at different times during radiation domination.}
	\label{SUBFIGURE LABEL 2}
 \end{subfigure}
 \caption{Power spectrum comparison for a massless test scalar field with past infinite vs finite duration dS inflation (dashed and bold lines, respectively) matched to a terminal stage of radiation domination. Finite duration inflation cures the would be IR divergences, leaving only UV divergences to attend to.}
 \label{fig:sfps}
 \end{figure}

We now use these results to find expressions for the coincident limits of the two point function and $\rho$ during the terminal stage of radiation domination. The two point correlation function coincidence limit can be expressed as 
\begin{eqnarray}
\nonumber	\lim_{x\to y}\langle\hat\phi(\tau,x)\hat\phi(\tau,y) \rangle &=& \frac{1}{2 \pi^2 a^2}  \int_0^{\infty} \frac{dk}{k} \: \frac{k^2}{2}\left[1+2 |\beta_k^{\rm R}|^2 \right. \\ \nonumber
&&\left. + \alpha^{\rm R}_k\beta^{\rm R*}_ke^{\frac{2 i k}{a_{\rm R}H}\left(2 - \frac{a}{a_{\rm R}}\right)} + \alpha^{\rm R*}_k\beta^{\rm R}_ke^{- \frac{2 ik}{a_{\rm R}H}\left(2 - \frac{a}{a_{\rm R}}\right)}\right]\\ \label{2pt00}  &=& \frac{1}{2 \pi^2 a^2}  \int_0^{\infty} \frac{dk}{k} \: \frac{k^2}{2}\left[1+2 |\beta_k^{\rm R}|_{\rm power}^2 + \{{\rm osc}\} \right]
\end{eqnarray}
where the first line has used $|\alpha_k^{\rm I} |^2 - |\beta_k^{\rm I} |^2=|\alpha_k^{\rm R} |^2 - |\beta_k^{\rm R} |^2 = 1$, and the second line splits the integrand into strictly power law contributions and oscillatory contributions. From Eqs. \ref{mcI2} and \ref{mcR}, we find that 
\begin{eqnarray}\label{bar}
	|\beta^{\rm R} _k|_{\rm power}^2=   \frac{a_{\rm R} ^4  H^4+ a_{\rm I} ^4 H^4}{4k^4}+\frac{a_{\rm I} ^4 a_{\rm R} ^4  H^8}{8k^8},
\end{eqnarray}
which nominally appears to have aggravated the IR divergence of the two point function. However, as we demonstrate in Appendix \ref{app:osc}, although the oscillatory terms can indeed be neglected for the purposes of extracting UV divergences, the oscillations freeze in the IR and exactly cancel the contributions from $|\beta^{\rm R} _k|_{\rm power}^2$. Specifically, one finds that 
\eqn{oscIR}{ \lim_{k\to 0} \{{\rm osc}\} = -\frac{a_{\rm R} ^4  H^4+ a_{\rm I} ^4 H^4}{2k^4} - \frac{a_{\rm I} ^4 a_{\rm R} ^4  H^8}{4k^8} - 1 + \frac{(3 a_{\rm I}^3 a_{\rm R} + 2 a (a_{\rm R}^3 -a_{\rm I}^3))^2}{9 a_{\rm I}^2 a_{\rm R}^6},}
which cancels the contributions from $1 + 2|\beta^{\rm R} _k|_{\rm power}^2$,  leaving only the UV divergence resulting from the final term of Eq. \ref{oscIR} to reckon with:
\begin{eqnarray}
	\label{2pt}
		\lim_{x\to y}\langle\hat\phi(\tau,x)\hat\phi(\tau,y) \rangle_{\rm div} &=& e^{-4N_{\rm tot}}\left(3 + \frac{2a}{a_{\rm R}}\left[e^{3\mathcal N_{\rm tot} - 1}\right]\right)^2\frac{1}{36 \pi^2 a^2}  \int_0^{\infty} dk \: k, 
\end{eqnarray}
where as before, $e^{-\mathcal N_{\rm tot}} = a_{\rm I}/a_{\rm R}$. Two things are to be immediately noted. Firstly, that the physical UV and IR scales $k_{\rm IR/ UV} = a_{\rm I/R} H$ associated with the beginning and end of inflation do not by themselves regulate the UV divergences. Instead, they merely parameterize the divergence through their ratio $k_{\rm IR}/k_{\rm UV} = e^{-\mathcal N_{\rm tot}} $ in the pre-factor above. Furthermore, the pre-factor itself diverges as $\mathcal N_{\rm tot} \to \infty$, indicative of the restoration of the pathologies of a past infinite dS space in that limit, where moreover, a logarithmic IR divergence reappears. This is perhaps best illustrated by Fig. \ref{fig:sfps}, where we plot all contributions of the power spectrum for finite duration inflation (identified through the logarithmic integrand of Eq. \ref{2pt00}) along with what would have resulted for a past-infinite dS cosmology matched to a terminal phase of radiation domination\footnote{The limit of a pre-inflationary phase of infinite duration inflation can straightforwardly be read off from inserting Eq. \ref{mcR} into Eq. \ref{mcI2} with $a_{\rm I} \equiv 0$.}. 
The UV divergence is unsurprisingly unchanged, and can be subtracted with a cosmological constant counterterm were one to impose a cutoff in physical momenta\footnote{Where factorizing the divergence in dimensional regularization as in the previous sub-section would instead see the UV contributions cancel among themselves.}.  
 
\begin{figure}
	\begin{subfigure}{.53\textwidth}
		\begin{flushleft}
		 \hspace{-1.3cm}
			\includegraphics[width=2.85in]{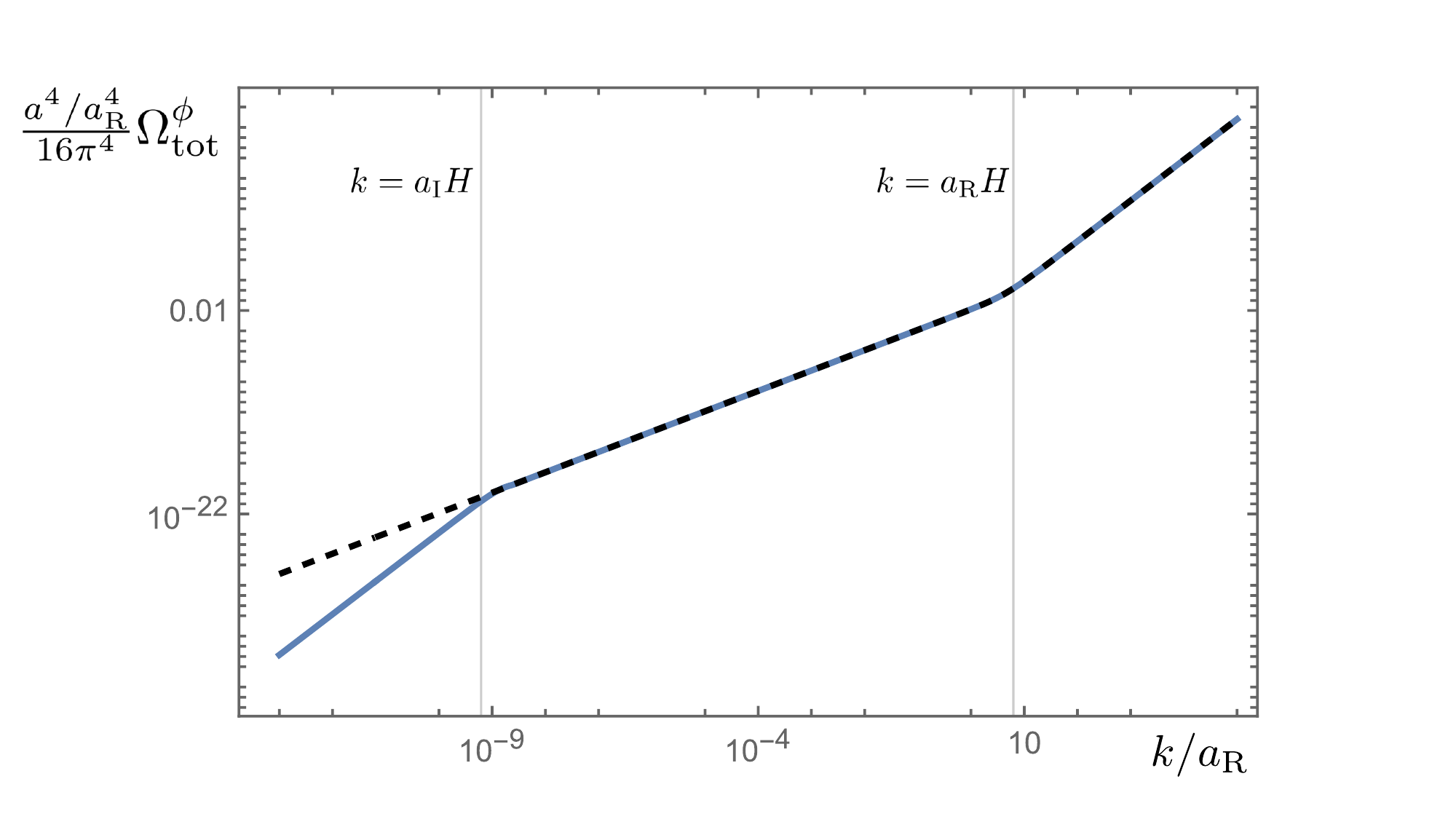}  
			\caption{}
			\label{SUBFIGURE LABEL 3}
		\end{flushleft}
	\end{subfigure}
	\begin{subfigure}{.53\textwidth}
	 \hspace{-1.3cm}
		\centering
		\includegraphics[width=2.85in]{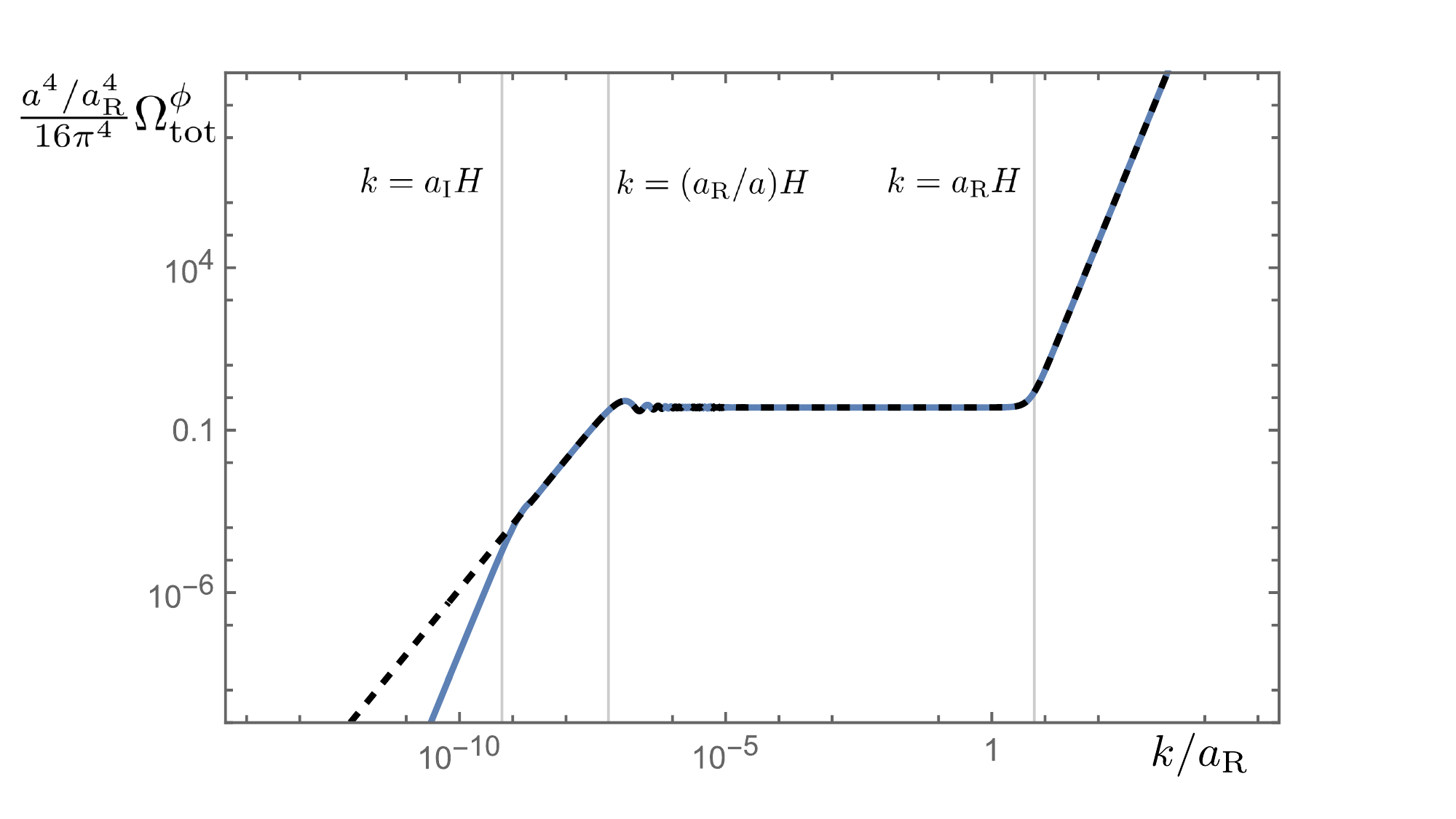}  
		\caption{}
		\label{SUBFIGURE LABEL 4}
	\end{subfigure}
	\begin{subfigure}{.53\textwidth}
	\begin{flushleft}
		 \hspace{-1.3cm}
		\includegraphics[width=2.85in]{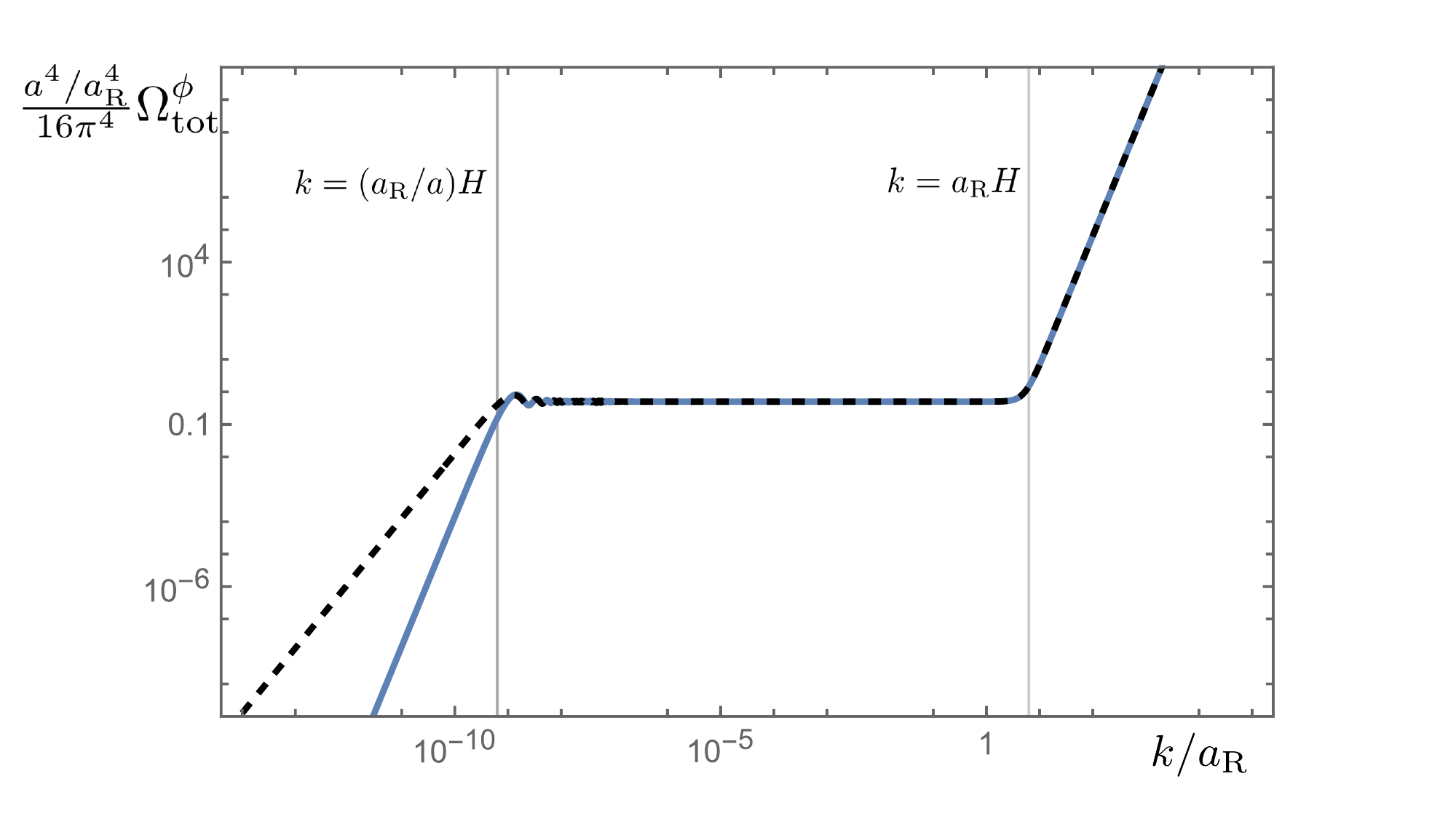}  
		\caption{}
		\label{SUBFIGURE LABEL 5}
	\end{flushleft}
 \end{subfigure}
 \begin{subfigure}{.53\textwidth}
 	 \hspace{-1.3cm}
	\centering
	\includegraphics[width=2.85in]{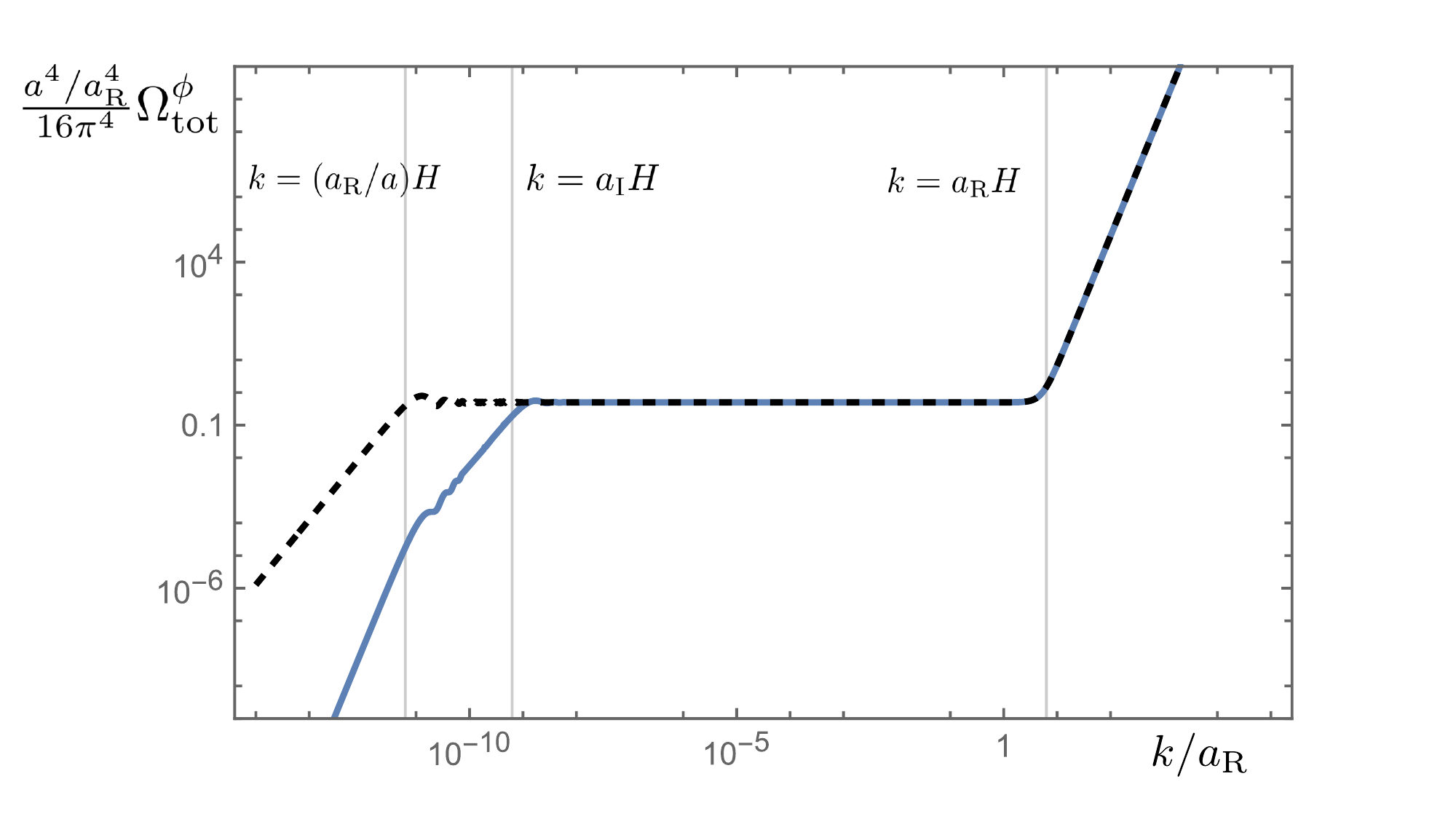}  
	\caption{}
	\label{SUBFIGURE LABEL 6}
 \end{subfigure}
	\caption{Power spectral energy density for a massless test scalar field comparing past infinite vs finite duration de Sitter inflation (dashed and bold lines, respectively) matched to a terminal stage of radiation domination. The panels are evaluated at subsequent times during radiation domination, with $a_{\rm I} = 10^{-10}a_{\rm R}$, and with $a/a_{\rm R} = 1, 10^{-8}, 10^{-10},$ and $10^{-12}$, respectively, in units where $H = 2\pi$.}
	\label{fig:sfom}
 \end{figure}

In computing the energy density $\rho$, we begin with
\eqn{r2pt4}{ \lim_{y\to x}\rho(\tau;x,y) = \frac{1}{4\pi^2 a^2}\int_0^\infty k^2\,dk\left[k^2\mathcal |\phi^{\rm RD}_k(\tau)|^2 + |\phi^{\rm RD \prime}_k (\tau)|^2\right]}
and use Eq. \ref{modRD2} to express the energy density during radiation domination as
\begin{eqnarray}
\nonumber 
\rho &=&  \frac{1}{8\pi^2 a^4}\int_{0}^\infty \,\frac{d k}{k}\left[ k^4\left(2 + \frac{a_{\rm R} ^4 H^2 }{a^2  k^2} \right)\left(1 + 2|\beta^{\rm R} _k|^2_{\rm power} \right) + \{{\rm osc}\}\right]
\\ \label{rhoRDscalar} &:=&  \int_{0}^\infty \,\frac{d k}{k} \left[\Omega^\phi_{\rm power} (k) + \Omega^\phi_{\rm osc}(k)\right]
\end{eqnarray}
where the integrated contribution of $\Omega^\phi_{\rm osc}$, defined as $\rho_{\rm osc}$ is given in Eq. \ref{roscpt1}, where it is explicitly shown not to contribute to any UV divergences. As was the case with the two point function, the contributions form Eq. \ref{bar} that nominally appear to aggravate IR divergences in the context of Eq. \ref{rhoRDscalar} are canceled by contributions from the oscillatory terms that freeze out in the IR. One can expand the sum of all contributions in the IR to find the IR safe scaling $\Omega^\phi_{\rm tot} \propto k^4$, to be compared to what would have been obtained had there been no pre-inflationary phase, where $\Omega^\phi_{\rm tot} \propto k^2$. We illustrate this behavior, and the processing of the logarithmic spectra power density as we go deeper into the radiation dominated regime in Fig. \ref{fig:sfom}, where the sub-horizon decay compensates for the $k^{2}$ scaling for the modes that exited the Hubble horizon during inflation to produce a scale invariant spectral density during radiation domination. We further note the additional long wavelength suppression of the spectral density for a pre-inflationary phase relative to past eternal dS. 
   
Returning to Eq. \ref{rhoRDscalar}, we see that upon inserting Eq. \ref{bar}, the only divergences that need to be regulated are given by the contributions
\begin{equation}\label{r2pt7}
	\rho_{\rm div} =\frac{1}{8\pi^4 a^4}\int_{-\infty}^\infty d^4 k  \left(1 + \frac{H^4a_{\rm R}^4 + H^4 a_{\rm I}^4}{2k^4}\right) +\frac{1}{8\pi^4 a^6}\int_{-\infty}^\infty d^4 k\frac{a_{\rm R}^4 H^2}{2 k^2},
\end{equation}  
where again we recast the logarithmic integration into the formally equivalent 4D Euclidean form as per Eq. \ref{sfd2}, and change integration variables to physical momenta $q = k/(a_{\rm R}H)$. We stress that all factors of $H$ in the expressions above correspond to the Hubble parameter during the intermediate phase on inflation, which is presumed dS and therefore fixed as per Eq. \ref{Hinfdef}. 

One can proceed from here as we did in the previous subsection in comparing the results from imposing cutoffs in physical momenta with that of dimensionally regularizing factorized scaleless integrals for the energy density, with the caveat that the counterterms are strictly speaking only to be identified only through a scheme that preserves diffeomorphism invariance as per the discussion at the end of the previous subsection. By imposing cutoffs in physical momenta: $k = a\Lambda_{\rm UV}$, we see that the first term in Eq. \ref{r2pt7} corresponds to a divergence of the form
\eqn{hcemt1}{\frac{1}{8\pi^4 a^4}\int_{-\infty}^\infty \,d^4 k = \frac{1}{4\pi^2 a^4}\int_{0}^{a \Lambda_{\rm UV}} \, k^3 d k = \frac{\Lambda_{\rm UV}^4}{16\pi^2},}
and the third term corresponds to a divergence of the form
\eqn{hcemt2}{\frac{1}{8\pi^4 a^6}\int_{-\infty}^\infty \,d^4 k \frac{a_{\rm R}^4 H^2}{2 k^2} = \frac{a_{\rm R} ^4 H^2}{8\pi^2 a^6}\int_{0}^{a \Lambda_{\rm UV}} \, k d k = \frac{\Lambda_{\rm UV}^2}{16\pi^2} \frac{H^2}{(a/a_{\rm R})^4},}
which would nominally be subtracted by a counterterm proportional to $R$ were we to insist on hard cutoff cutoff regularization. The former, when varied with respect to the metric yields the contribution $R_{\mu\nu} \sim 1/a^4$ and thus corresponds to a renormalization of Newton's constant. Finally, the second term in Eq. \ref{r2pt7} result in a divergence of the form
\begin{eqnarray} \nonumber 
\frac{1}{8\pi^4 a^4}\int_{-\infty}^\infty \,d^4 k \: \left(\frac{a_{\rm R}^4  H^4+ a_{\rm I}^4 H^4}{2k^4}\right) &=& \frac{1}{8\pi^2 a^4}\int_{a \Lambda_{\rm IR}}^{a \Lambda_{\rm UV}} \,\frac{d k}{k} \left(a_{\rm R} ^4  H^4+ a_{\rm I}^4 H^4\right) \\ \label{corho}
&=& \frac{H^4(1 + e^{-4 \mathcal N_{\rm tot}})}{8\pi^2 (a/a_{\rm R})^4}\log \frac{\Lambda_{\rm UV}}{\Lambda_{\rm IR}},
\end{eqnarray} 
where the scale $\Lambda_{\rm IR}$ does not have the same significance as before given the IR safe behavior of the spectral density, and where the above also corresponds to a renormalization of Newton's constant. We note that the scales corresponding to the beginning and end of inflation appear in the coefficient of the logarithm, whereas the UV scale corresponding to the unknown completion the theory appears inside the logarithm.

Were we to now dimensionally regularize the divergences as advised, we can perform a similar factorization of the scaleless integrals as in the previous subsection. The nominally power law divergent parts can be isolated as
\eqn{rUV}{\begin{aligned}
\rho^{\rm UV}_{\rm div} &= \frac{1}{8\pi^4 a^4}\int_{-\infty}^\infty \,d^4 k\left(1 + \frac{a_{\rm R}^4 H^2}{2 a^2 k^2}\right) \\
	 &=\frac{1}{8\pi^4 a^4}\int_{-\infty}^\infty \,d^4 k\left[\frac{k^2 + m^2}{(k^2 + m^2)}  + \frac{a_{\rm R}^4 H^2}{2 a^2 }  \left(\frac{1}{(k^2 + m^2)} + \frac{m^2}{k^2(k^2 + m^2)} \right)\right],
\end{aligned}}
where we see that power law UV divergences cancel among themselves, and so do not necessitate any counterterms. On the other hand, the UV divergent logarithmic term can be isolated and factorized as
\begin{eqnarray}
 \nonumber 	\rho^{\rm log}_{\rm div} &=& \frac{1}{8\pi^4 a^4}\int_{-\infty}^\infty \,d^4 k \: \left(\frac{a_{\rm R}^4  H^4+ a_{\rm I}^4 H^4}{2k^4}\right) \\
	\label{rLOG}&=&\frac{1}{8\pi^4 a^4}\int_{-\infty}^\infty \,d^4 k \left( \frac{a_{\rm R}^4 H^4+a_{\rm I}^4 H^4}{2  }  \right) \left(\frac{1}{k^2(k^2 + m^2)} + \frac{m^2}{k^4(k^2 + m^2)}  \right).
\end{eqnarray} 
By isolating the logarithmic UV poles, we find  
\eqn{rholog}{\rho^{\rm log}_{\rm div}= \frac{H^4(1 + e^{-4\mathcal N_{\rm tot}})}{8\pi^2 (a/a_{\rm R})^4 } \left[  \frac{1}{\delta_{\rm UV}} + 1 - \gamma_E +  \log\left(\frac{\mu}{H}\right) \right] }
from which we read off an identical coefficient as computed in Eq. \ref{corho}\footnote{One can also show that the analogous computation for the pressure component also necessitates a counterterm corresponding to a renormalization of $G_{N}$ (i.e. resulting in a divergent contribution proportional to the spatial components of the Ricci tensor when varied with respect to the background metric) resulting in a renormalized stress tensor that is traceless on a radiation domination background. We do so explicitly for the case of tensor modes in Appendix \ref{app:emt}.}. For completeness, we show in Appendix \ref{app:ps} that the coefficient of the logarithmic divergence obtained from point split regularization in the FRLW slicing is also identical, with the same power law divergences identified in physical cutoff regularization up to the expected scheme dependence in the coefficients of the latter. 

To summarize our findings in this subsection: the scales corresponding to the beginning and end of inflation must in principle be separated from any UV and IR scales parameterizing the unknown completion of our theory and well defined observable quantities. Although physical quantities cannot depend on the latter, they may certainly depend on the former. We found that all IR divergences encountered in the simplified settings we worked in were regulated by the existence of a pre-inflationary phase, and therefore an artifact of the approximation of a past infinite dS phase. A highly pertinent question is whether this persists once we move to incorporate higher point interactions and couplings with other test fields. Unsurprisingly, UV divergences persist for backgrounds corresponding to finite duration inflation, and the scale corresponding to the end of inflation parameterizes their coefficients rather than regulating them. Upon subtraction of these divergences with local counterterms and the imposition of renormalization conditions, one is able to draw meaningful physical conclusions. However, the two point function and stress tensor of a test scalar field are of relatively academic interest given the assumed negligibility of the scalar field background, and therefore inability to fix renormalization conditions from observation. The analogous set of questions for vacuum tensor perturbations presents a far more interesting application given the assumption of an evolving background gravitational field (that of FRLW cosmology), perturbations around which represent gravitational waves, which is where we turn our attention towards next.   

\section{Gravitational waves}

Before delving into the substance of this section, it is informative to contrast the results obtained in the previous section to analogous expressions typically found in the literature in the context of gravitational waves (see e.g. \cite{Meerburg:2015zua, Maggiore2, Gasperini, Smith, Boyle, Lizarraga, Henrot, Cabass, Liu, Pagano, Li, Benetti, Berbig, Giare}), where imposing hard cutoffs on the energy density of gravitational waves results in expressions of the form: 
\eqn{MH}{ \rho_{\mathrm{GW}} \simeq \frac{ A_{t}}{32 \pi  G_{ N}}\left(\frac{k_{\mathrm{UV}}}{k_*}\right)^{n_{ t}} \frac{1}{2 n_{ t}} \frac{1}{a^4} \propto \frac{1}{a^4}\left[\frac{1}{n_{ t}} + \log\frac{k_{\rm  UV}}{k_{*}} \right],}
where $k_*$ is some reference IR scale, and the approximation is only valid when $n_t \to 0$ \cite{Meerburg:2015zua}. We deconstruct and rederive the energy density for vacuum tensor perturbations in what follows, but before doing so, it is useful to compare what one would have obtained in retracing the steps leading to Eq. \ref{MH} for a massless test scalar field, which would have resulted in the expression 
\eqn{}{ \rho \simeq \lim_{n_{s} \to 1}\frac{A_{s}}{32 \pi  G_{ N}}\left(\frac{k_{\mathrm{UV}}}{k_*}\right)^{n_{\rm s} - 1} \frac{1}{2 (n_{ s} - 1)} \frac{1}{a^4} \propto \frac{1}{a^4}\left[\frac{1}{n_{ s} - 1} + \log\frac{k_{\rm UV}}{k_{*}} \right],}
which can directly be compared to the divergences computed in Eqs. \ref{hcemt1} - \ref{corho} from hard cutoffs in physical momenta and Eq. \ref{rholog} from dimensional regularization, all of which are understood to be intermediate expressions that are to be subtracted and renormalized. 

An immediate reservation one might express for expressions such as Eq. \ref{MH} is the appearance of cutoffs in what should be a physical result. Although one might argue that the effects of UV and IR modes are negligible relative to what can be extracted from Eq. \ref{MH}, its form nevertheless suggests that the expression above is an intermediate result on the way to computing a physical energy density. Moreover, one might be concerned in applying the formula Eq. \ref{MH} to the case of blue tilted spectra (i.e. for positive $n_t \sim  \mathcal O(1)$, so that $\rho_{\rm GW} \sim (k_{\rm UV}/k_*)^{n_t}$ as derived in e.g. \cite{Meerburg:2015zua}), that some part of this expression is nothing other than the Fourier transform of a UV divergence that gets subtracted in the usual way. The particular case of $n_t = 3$ is particularly interesting in this context, as part of it is exactly mimicked by the familiar quadratic UV divergence from the perspective of cutoff regularization, necessitating greater care in the interpretation of such power spectra. In spite of this caution, it is both practical and straightforward to follow through this process to the end to ensure that one is computing well defined physical observables. In the specific context of vacuum tensor perturbations, however, additional subtleties must be accounted for. 

The formula for the stress energy tensor of gravitational waves was first derived by Isaacson in the context astrophysically sourced gravitational waves \cite{Isaacson} (see also \cite{Maggiore, Maccallum:1973gf}), which is nevertheless extensively referenced in computations of the stress energy tensor of cosmological gravitational wave backgrounds\footnote{e.g. leading to Eq. \ref{MH}. See also \cite{Caprini:2018mtu} and references therein.}, and is given by\footnote{Implicit in the expectation value of Eq. \ref{Rho gw literature 0 main} is an additional (Brill-Hartle) spatial and temporal averaging prescription. We presume this as implicit when discussing the Isaacson form of the stress tensor in what follows.} 
\begin{equation}\label{Rho gw literature 0 main}
	\begin{aligned}
		\rho^{\rm Isc}_{\rm gw} = \frac{1}{32 \pi a^2 G_N} \left\langle h^{\prime}_{i j}(\tau, k) h^{\prime i j}(\tau, k)\right\rangle.
	\end{aligned}
\end{equation}
However, retracing the steps used in its derivation, it becomes immediately clear that one needs to proceed with extra care for applications in a cosmological context, in particular when considering questions of renormalized stress tensors in prescriptions that involve integrations over all momenta\footnote{A feature that can be bypassed entirely in the context of Hadamard regularization \cite{Allen:1987bn, NP}.}. This is because the Isaacson form presumes a prior scale separation between fast and slow frequencies defined relative to the curvature scale of the background (a criteria that is time dependent), and repeatedly relies on this scale separation along with the implicit averaging prescription to facilitate various approximations resulting in the simplified final form Eq. \ref{Rho gw literature 0 main}. In the context of an expanding spacetime, with modes of observational relevance that might have crossed any a priori defined scale more than once over cosmic evolution, undoing the steps that relied on these approximations is warranted. We do so in Appendix \ref{app:gw}, retracing the derivation of the stress tensor for gravitational waves, finding the resulting expression for the energy density\footnote{From now on, even if we drop the hat in the notation, $h_{ij}$ is an operator as defined in Eq. \ref{eq:gamma}.}:
\begin{equation}\label{eq:rho 1 main}
	\begin{aligned}
	\rho^{\rm Imp}_{\rm gw}=\frac{1}{64 \pi  G_N }\left\langle \frac{1}{a^2} \bigg[ h_{ij}^{\prime} h^{\prime  ij}- 3 \partial_k h_{ij} \partial^k h^{ ij} -4  h_{ij} \partial_k\partial^k h^{ ij}+2 \partial_k h_{ij} \partial^j h^{i k}  +8\mathcal{H} h_{j}{}^{ i} h^\prime_{i}{}^{ j}  \bigg] \right\rangle
	 	\end{aligned}
\end{equation}
where $\mathcal H := a'/a$. The derivation is straightforward, and follows from the same starting point as the derivation leading to Eq. \ref{Rho gw literature 0 main}, but with any steps that invoked a time averaging prescription or frequency separation left undone (we elaborate on the details in Appendix \ref{app:gw}). 

It is worth stressing that Eq. \ref{eq:rho 1 main} is the temporal component of a covariantly conserved tensor, and is under no obligation to conserved in isolation. This is a corollary of the fact that on FRLW backgrounds, a locally conserved energy cannot be defined. Moreover, it is also under no obligation to be positive definite. In fact, were one to insist of constructing the spectral density associated with the vacuum expectation value of Eq. \ref{eq:rho 1 main}, one can show that it crosses zero at a comoving scale corresponding to the horizon scale at any given time, consistent with the operational ambiguity of associating a background/fluctuation split for wavelengths commensurate with the background curvature. However, none of this is of any operational concern for our purposes, as we stress that Eq. \ref{eq:rho 1 main} is to be viewed as the components of a covariantly conserved tensor corresponding to a massless spin two excitation, whose role in renormalization of background quantities is well understood in the covariant context, but obscured and widely conflated for a physical contribution to the number of relativistic species in foliation specific computations as we elaborate upon next. The specific computation we are interested in performing is the renormalization of the graviton stress tensor on a background corresponding to radiation domination preceded by a finite period of inflation. In doing so, we will compare and contrast the results obtained from both the Isaacson and improved forms of the stress tensor, before revisiting the question of $N_{\rm eff}$ bounds from vacuum tensor modes and concluding.

\subsection{Divergences -- vacuum tensor modes and finite duration inflation}\label{subsec:gwfi} 
Working in transverse-traceless (TT) gauge (see Appendix \ref{app:gw} for the relevant details and caveats), we expand the tensor perturbations as 
\begin{equation}\label{eq:gamma}
	\hat{h}_{ij}(\tau,x) = \sum_{r=+,x}\int \frac{d^{3}k}{M_{\rm pl}  \left(2 \pi \right)^{3 }}  e^{ix \cdot k } \left[\epsilon_{ij}^r (k) \hat{a}_{k} \gamma_k(\tau) + \epsilon_{ij}^{r*} (-k)  \hat{a}_{-k}^{ \dag} \gamma^*_k (\tau) \right]
\end{equation}
with polarization tensors normalized as  $\epsilon_{ij}^{s} \epsilon_{ij}^{r*}=4 \delta^{rs}$, $k^i \epsilon_{ij}^{r}=0$. The normalizations are chosen so that the relevant mode functions during inflation and the pre and post-inflationary phases of radiation domination are given for each polarization just as for massless minimally coupled scalars\footnote{The EOM for $h^{ j}_{i} $ in TT gauge can be derived from the first order expansion of Einstein equation: $ \hat{h}^{\prime \prime j} _{i} + 2 \mathcal{H} \hat{h}^{ \prime j} _{i} - \partial_k^2  \hat{h}^{  j} _{i}=0.$}. These are:
\begin{eqnarray} \nonumber
	\gamma^{\rm PI}_k (\tau)&=& \frac{1}{a}\frac{1}{  \sqrt{2k}}e^{-i k \tau_{\rm I} \left( 2 - \frac{a}{a_{\rm R}} e^\mathcal N_{\rm tot}\right)} \\
	\gamma'^{\rm PI}_k (\tau)&=& -\frac{1}{a^2 \sqrt{2k}} \frac{a_{\rm I} }{\tau_{\rm I} } e^{-i k \tau_{\rm I} \left( 2 - \frac{a}{a_{\rm R} } e^\mathcal N_{\rm tot}\right) }\left(-1 + i k \tau_{\rm I}  \frac{a}{a_{\rm I} } \right),
\end{eqnarray}
during the pre-inflationary phase,
\begin{eqnarray}  \nonumber
	\gamma_k^{\rm I} (\tau) &=& \frac{H}{ \sqrt{2k^3}}\left[\alpha_k^{\rm I}  e^{i \frac{k}{aH}}\left(1 - \frac{i k}{a H}\right) +  \beta_k^{\rm I}  e^{-i \frac{k}{aH}}\left(1 + \frac{i k}{a H}\right)\right] \\
	\gamma'^{\rm I}_k (\tau) &=& -\frac{H}{\sqrt{2k^3}}\left[\alpha_k^{\rm I}  e^{i \frac{k}{aH}}\frac{k^2}{a H} +  \beta_k^{\rm I}  e^{-i \frac{k}{aH}}\frac{k^2}{aH}\right],
\end{eqnarray}
during inflation, and
\eqn{modRD2gw}{\begin{aligned}  
\gamma^{\rm RD}_k (\tau)&= \frac{1}{a}\frac{1}{ \sqrt{2k}}\left[\alpha_k^{\rm R} e^{-i k \tau_{\rm R} \left( 2- \frac{a}{a_{\rm R}}\right)} + \beta_k^{\rm R} e^{i k \tau_{\rm R} \left( 2- \frac{a}{a_{\rm R}}\right)}  \right] \\
	\gamma'^{\rm RD}_k (\tau) &= \frac{a_{\rm R}}{a^2\tau_{\rm R}}\frac{1}{ \sqrt{2k}}\left[\alpha_k^{\rm R} e^{-i k \tau_{\rm R} \left( 2- \frac{a}{a_{\rm R}}\right)}\left(1 - i  \frac{a k \tau_R}{a_{\rm R}}\right) + \beta_k^{\rm R} e^{i k \tau_{\rm R} \left( 2- \frac{a}{a_{\rm R}}\right)}\left(1 +i  \frac{k \tau_{\rm R} a}{a_{\rm R}}\right)  \right]
\end{aligned} }
during the terminal radiation dominated phase. The corresponding Bogoliubov coefficients are the same as those given in Eq. \ref{mcI2} and Eq. \ref{mcR}. 

An immediate corollary of the above is that the two point correlation function of each graviton polarization is identical to that of a massless, minimally coupled scalar. Therefore, as illustrated in Figs. \ref{SUBFIGURE LABEL 1} and \ref{SUBFIGURE LABEL 2}, one finds that the IR divergences exhibited on a past infinite dS background are also cured for the graviton two point function on a background corresponding to finite duration inflation. 

The energy density of gravitational waves presuming the Isaacson form of the stress tensor Eq. \ref{Rho gw literature 0 main} results in (with the reduced Planck mass defined as $M^2_{\rm pl} = \frac{1}{8 \pi G_N}$):
\begin{eqnarray}\label{rhogw1}
	\rho_{\rm gw}^{\rm Isc}&=& \lim_{y\to x}\rho_{\rm gw}(\tau;x,y)= \frac{1}{ \pi^2 a^2} \int_0^{\infty}dk \ k^2 \: \left[\gamma^{' \rm RD}_k\gamma^{' \rm RD *}_k \right]
\end{eqnarray}
as the tensor mode counterpart of Eq. \ref{r2pt4}, after having summed the contributions from the two independent polarizations. 
 \begin{figure}
	\begin{subfigure}{.53\textwidth}
		\begin{flushleft}
		\hspace{-1.3cm}
			\includegraphics[width=3.05in]{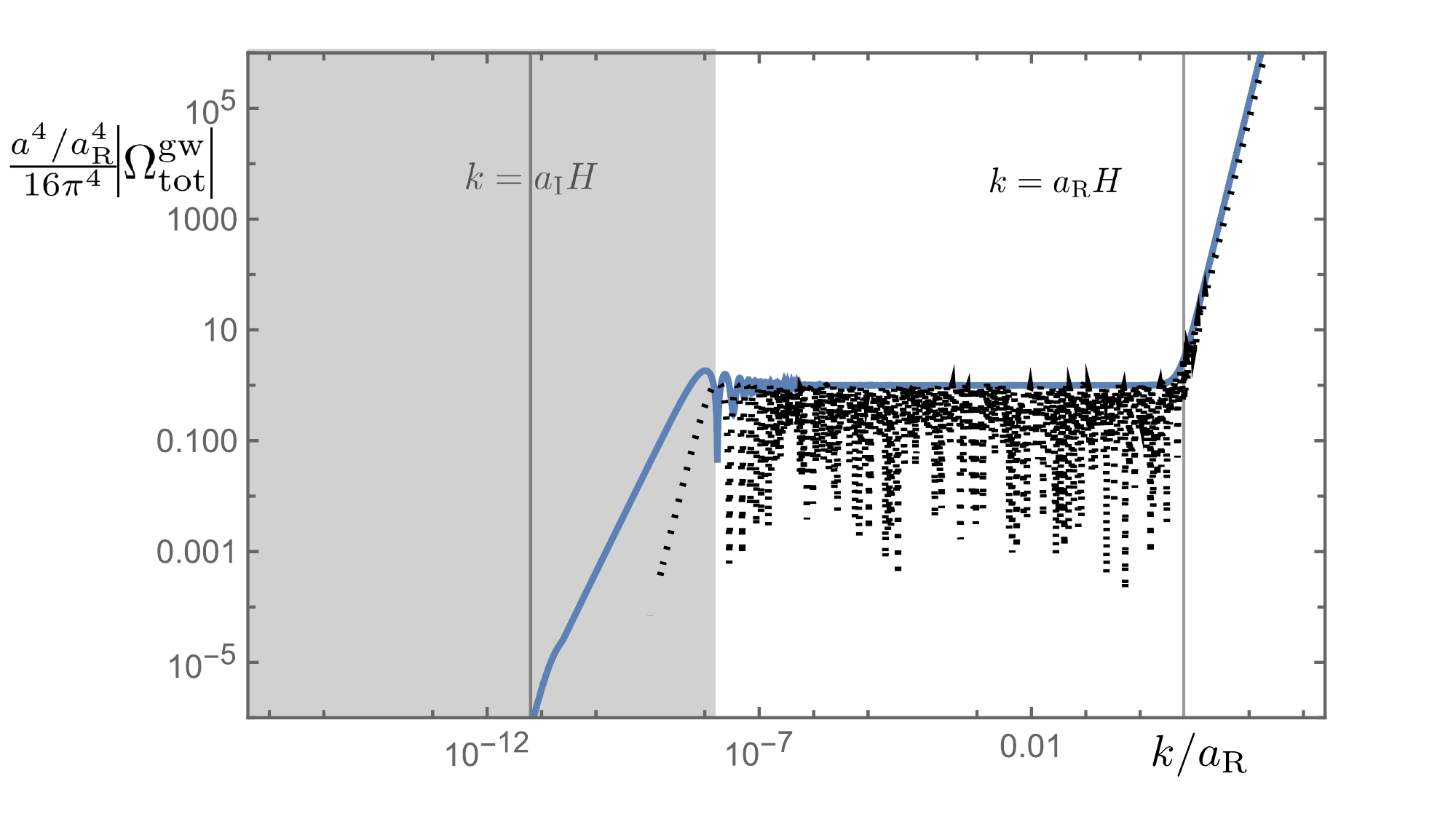}  
		\end{flushleft}
	\end{subfigure}
	\begin{subfigure}{.53\textwidth}
			\hspace{-1.3cm}
		\centering
		\includegraphics[width=3.05in]{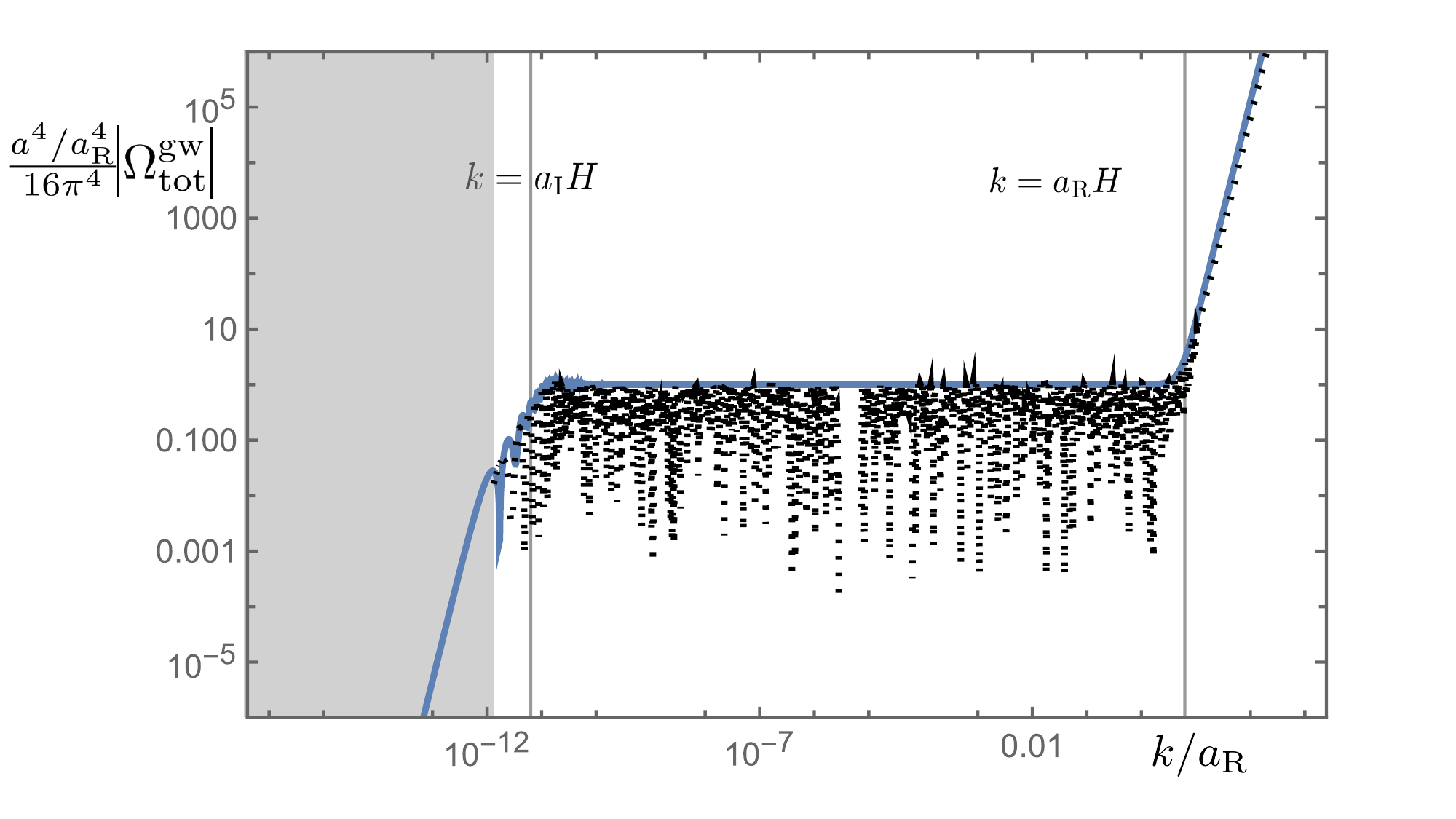}  
	\end{subfigure}
	\caption{Power spectral density of the Isaacson stress tensor (dashed lines) evaluated at $a = 10^9 a_{\rm R}$ (left) and $a = 10^{13} a_{\rm R}$ (right), with $a_{\rm I} = 10^{-12} a_{\rm R}$ in units where $H$ is set to $2\pi$. The blue line is the comparison to the spectral density of the improved stress tensor. The gray shaded regions correspond to super-horizon scales, which are outside the domain of validity of the Isaacson stress tensor (where we note that the oscillations would not appear in its time averaged form), and also where the spectral density for the improved stress tensor nominally becomes negative (cf. discussion below Eq. \ref{eq:rho 1 main}).}
	\label{fig:gwIsc}
 \end{figure}
Using the mode functions specified by Eqs. \ref{modRD2gw}, the relations \ref{mcR}, and tracing through the steps of the previous section, we find 
\begin{eqnarray} \nonumber
	\rho_{\rm gw}^{\rm Isc}&=&\frac{1}{2  \pi^2 a^4}  \int_0^{\infty} \frac{d k}{k} \: k^4 \left[ \left(1 + \frac{a_{\rm R}^4 H^2 }{ a^2  k^2} \right)\left(1 + 2|\beta^{\rm R}_k|^2_{\rm power} \right)  + \{{\rm osc}\}\right]
	\\ \label{rhogw2}  &:=&   \int_{0}^\infty \,\frac{d k}{k} \left[\Omega^{\rm gw}_{\rm power} (k,\tau) + \Omega^{\rm gw}_{\rm osc}(k,\tau)\right],
\end{eqnarray}
where, as we did for the scalar case, we similarly separate the spectral density into oscillatory and power law contributions, and all contributions to $\rho_{\rm gw}$ can be found in Eq. \ref{rhogwfull}. Even though the Isaacson form of the stress tensor is not strictly valid when incorporating wavelengths greater than the Hubble scale at any given time, na\"ively persisting with it for all wavelengths would show an IR regular spectral density $\Omega^{\rm gw}_{\rm total} \propto k^2$, with oscillatory contributions contributions giving negligible contributions in the UV. Consequently, one only has the resulting UV divergent contributions to regulate:
\begin{eqnarray}\label{rhogw3}
	\rho^{\rm Isc}_{\rm gw, div} &=&\frac{1}{4  \pi^4 a^4}   \int_{-\infty}^\infty \,d^4 k \: \left(1+\frac{a_{\rm R}^4  H^4+ a_{\rm I}^4 H^4}{2k^4}+\frac{ a_{\rm R}^4  H^2}{a^2 k^2} \right),
\end{eqnarray}
which up to overall factors, is formally identical to the contributions Eq. \ref{r2pt7}, and whose subtraction proceeds along the same lines. 

Were we to consider the improved stress tensor Eq. \ref{eq:rho 1 main} (which does not invoke a scale separation or time averaging prescription), one would obtain
\eqn{}{ \rho_{\rm gw}^{\rm Imp} =\frac{1}{2 \pi^2a^2} \int_0^{\infty}dk \ k^2 \: \left[\gamma^{' \rm RD}_k\gamma^{' \rm RD *}_k+ k^2  \gamma^{\rm RD}_k\gamma^{\rm RD *}_k  + 4 \mathcal{H} \left( \gamma^{' \rm RD}_k\gamma^{\rm RD *}_k+ \gamma^{\rm RD}_k\gamma^{' \rm RD *}_k \right) \right] }
as the counterpart of Eq. \ref{rhogw1} that is now valid for for all wavelengths. Using the mode functions and normalizations specified above, one finds 
\begin{eqnarray} \nonumber
		\rho_{\rm gw}^{\rm Imp}&=&\frac{1}{2  \pi^2 a^4}  \int_0^{\infty} \frac{d k}{k} \: k^4 \left[ \left(1 - \frac{7 a_{\rm R}^4 H^2 }{2 a^2  k^2} \right)\left(1 + 2|\beta^{\rm R}_k|^2_{\rm power} \right)  + \{{\rm osc}\}\right],
		\\ \label{ImpSD} &:=&  \int_{0}^\infty \,\frac{d k}{k} \left[\Omega^{\rm gw}_{\rm power} (k,\tau) + \Omega^{\rm gw}_{\rm osc}(k,\tau)\right]
\end{eqnarray}
where we can again separate the spectral density\footnote{We stress that the spectral density as defined in Eq. \ref{ImpSD} is to be viewed as only a calculational definition for the purposes of comparison to the literature, and not be to viewed as the power spectral density of vacuum fluctuations without additional caveats.} into oscillatory and power law contributions, where all contributions to $\rho^{\rm Imp}_{\rm gw}$ can be found in Eq. \ref{rhoimproved}. 

One again finds that the spectral density summing all contributions in the IR goes as $\Omega^{\rm gw}_{\rm total} \propto k^2$, albeit with a negative overall coefficient (cf. the discussion below Eq. \ref{eq:rho 1 main}), with the following divergent contributions in the UV that necessitate subtraction 
	\begin{eqnarray}\label{}
		\rho_{\rm gw, div}^{\rm Imp} &=&\frac{1}{4  \pi^4a^4}  \int_{-\infty}^\infty \,d^4 k \: \left(1+\frac{a_{\rm R}^4  H^4+ a_{\rm I}^4 H^4}{2k^4}-\frac{7\, a_{\rm R}^4  H^2}{2a^2 k^2} \right).
	\end{eqnarray}

For illustrative purposes, we plot the power spectral densities of the (time unaveraged) Isaacson and improved forms of the stress tensors in Fig. \ref{fig:gwIsc}. For sub-horizon modes, where a positive spectral density results, the improved stress tensor evaluated at any given time does not exhibit oscillations\footnote{The oscillatory behavior of the time unaveraged Isaacson stress tensor can be understood from the fact that the magnitude of the time derivative squared of a linearized gravitational wave is by itself not a constant of the linearized equations of motion (which requires restoring the spatial derivative contributions), unlike for the improved form.}.
In proceeding to regularizing divergences, we can identity the various UV divergences is different schemes, finding identical coefficients for the log divergences in all of them. By imposing cutoffs in physical momenta ($k = a\Lambda_{\rm UV}$) on the identifiably UV-divergent terms above, we obtain
	\begin{eqnarray}\label{}
		\frac{1}{2  \pi^2 a^4} \int_{0}^{a \Lambda_{\rm UV}} d k  \: k^3 -\frac{7}{2} \frac{ a_{\rm R}^4 H^2}{2  \pi^2 a^6}  \int_{0}^{a \Lambda_{\rm UV}} d k  \: k =\frac{\Lambda_{\rm UV}^4}{8\pi^2} - \frac{7 H^2 \Lambda_{\rm UV}^2}{8  \pi^2 (a/a_{\rm R})^4}
	\end{eqnarray}
and
\begin{eqnarray}\label{}
		\frac{1}{4\pi^2 a^4}\int_{a \Lambda_{\rm IR}}^{a \Lambda_{\rm UV}} \,\frac{d k}{k} \left(a_{\rm R} ^4  H^4+ a_{\rm I}^4 H^4\right)=\frac{H^4\left(1 + e^{-4\mathcal N_{\rm tot}}\right)}{4 \pi^2 (a/a_R)^4}\log \frac{\Lambda_{\rm UV}}{\Lambda_{\rm IR}}.
	\end{eqnarray}
If instead we were to now isolate the power law UV divergent parts and dimensionally regularize, we would obtain 
\eqn{}{\begin{aligned}
\rho^{\rm Imp}_{\rm gw, div} \supset& \frac{1}{4  \pi^4a^4}  \int_{-\infty}^\infty \,d^4 k -\frac{7 a_{\rm R}^2}{4  \pi^4 a^6}  \int_{-\infty}^\infty \,d^4 k \: \left(\frac{a_{\rm R}^2  H^2}{2 k^2}\right) \\
		 =&\frac{1}{4\pi^4 a^4}   \int_{-\infty}^\infty \,d^4 k\left[\frac{k^2+ m^2}{(k^2 + m^2)} - \frac{7}{2} \frac{a_{\rm R}^4 H^2}{2 a^2 }  \left(\frac{1}{(k^2 + m^2)} + \frac{m^2}{k^2(k^2 + m^2)} \right) \right]
\end{aligned}}
which, as for the dimensionally regularized scalar case Eq. \ref{rUV}, has the divergent contributions canceling among themselves and not necessitating any counterterms. The logarithmic divergence can be expressed as
	\begin{eqnarray} \nonumber
		\rho^{\rm Imp}_{\rm gw, div}&\supset&\frac{1}{4  \pi^4 a^4 }   \int_{-\infty}^\infty \,d^4 k \: \left(\frac{a_{\rm R}^4  H^4+ a_{\rm I}^4 H^4}{2k^4} \right) \\
		&=  &   \frac{ H^4\left(1 + e^{-4\mathcal N_{\rm tot}}\right)}{4\pi^2 (a/a_{\rm R})^4 }   \left[  \frac{1}{\delta_{\rm UV}} + 1 - \gamma_E + \log\left(\frac{\mu}{H}\right) \right].
	\end{eqnarray}
Lastly, one can point split regularize Eq. \ref{rhogw3} by commuting the coincident limit with the momentum integrations, so that one has to evaluate
\begin{eqnarray}\label{}
		\rho_{\rm gw}^{\rm Imp} &=&\frac{1}{4  \pi^2 a^4 r}    \int_{0}^\infty \,k^2 d k\,\sin kr\:  \left[ \left(1 - \frac{7 a_{\rm R}^4 H^2 }{2 a^2 k^2} \right)\left(1 + 2|\beta^{\rm R}_k|^2_{\rm power} \right) + \{{\rm osc}\} \right],
	\end{eqnarray}
which can be shown to result in the UV divergent contributions
	\eqn{r2reg3}{\rho^{\rm Imp}_{\rm gw, div} =  \lim_{\sigma \to 0}  \left\{ -\frac{1}{\pi^2\sigma^4} -\frac{7 H^2}{4\pi^2\sigma^2(a/a_{\rm R}^4)} + \frac{  H^4\left(1 + e^{-4\mathcal N_{\rm tot}}\right) }{4 \pi^2 (a/a_{\rm R})^4}\left[1 - \gamma_E - \log{k_p\sigma} \right] \right\}, }
	where as before $\sigma = a r$ and $k_p := k/a$ are physical quantities. We mote that the coefficient of the logarithmic divergence is identical, with the minus sign reflecting the fact that $\sigma \to 0$ inside the point split logarithm, whereas $\Lambda_{\rm UV} \to \infty$ with hard cutoff regularization, the results of which can be collected and summarized as
\eqn{}{	\rho^{\rm Imp}_{\rm gw, div} =  \lim_{\Lambda_{\rm UV} \to \infty}  \left\{ \frac{1}{2  \pi^2 } \frac{\Lambda_{\rm UV}^4}{4}  -\frac{7}{2} \frac{ a_{\rm R}^4 H^2}{2  \pi^2 a^4} \frac{\Lambda_{\rm UV}^2}{2} +\frac{  H^4\left(1 + e^{-4\mathcal N_{\rm tot}}\right) }{4\pi^2 (a/a_{\rm R})^4}\log \frac{\Lambda_{\rm UV}}{\Lambda_{\rm IR}} \right\}, ~~ {\rm (cutoff)}}
and also compared to the answer obtained via dimensional regularization
\eqn{drimp0}{
		\rho^{\rm Imp}_{\rm gw, div} =  \lim_{\delta_{\rm UV} \to 0}  \left\{ \frac{H^4\left(1 + e^{-4\mathcal N_{\rm tot}}\right)}{4\pi^2 (a/a_R)^4}  \left[  \frac{1}{\delta_{\rm UV}} + 1 - \gamma_E + \log\left(\frac{\mu}{H}\right) \right] \right\}. ~~~~~~~~ {\rm (dim-reg)} }
After having subtracting divergences with the appropriate counterterms, a finite remainder remains that needs to be fixed by the imposition of renormalization conditions by observations at a given scale. This in order to have a predictive theory with which we can calculate the results for any subsequent observations. As cautioned in the previous section, one may not be able to do this reliably unless one uses regularization schemes that preserve the symmetries of the background so that the required counterterms can be consistently identified (see Appendix A for more details). For this reason, we stick to dimensional regularization in what follows, although covariant point splitting methods also offer a practical alternative\footnote{The point split regularization elaborated upon here is contingent on the FRLW slicing, which permits a Fourier expansion, and so cannot be considered fully covariant.} \cite{Birrell:1982ix, Adler:1976jx, Christensen:1978yd, Davies:1977ze, Christensen:1976vb, Bunch:1978aq, Bunch:1978yw}. 

In order to understand which counterterms we need to add in order to reabsorb the divergences in dimensional regularization, we need only to look at the scale factor dependence of the divergences in Eq. \ref{drimp}. However, we immediately notice that the divergences we have computed do not necessitate higher derivative counterterms beyond those already in the Einstein Hilbert action.
In order to understand why this is, we trace through a treatment that uses adiabatic regularization to renormalize the stress tensor of a non-minimally coupled test scalar field \cite{Bunch:1980vc}, where mode functions are adiabatically expanded in order to regularize divergences and identify counterterms. Higher orders in the adiabatic expansion necessitate successively higher order counterterms. The adiabatic solution of order zero is regularized by a cosmological constant counterterm, the second order solution by a curvature counterterm, with curvature squared counterterms needed to renormalize divergences that appear only at fourth order in the adiabatic expansion. In our example, the simplicity of the example of a minimally coupled massless field permits the exact computation of the mode functions via a matching calculation, obviating the need for any adiabatic approximation. Terms that would correspond to higher order adiabatic corrections would only be necessitated were one to consider massive fields, couplings to other massive scalar fields, or by incorporating the effects of loops and higher order gravitational and matter coupling non-linearities (see \cite{Landete:2013lpa, Glavan:2015cut, Ferreiro:2022ibf, Maranon-Gonzalez:2023efu, Animali:2022lig} for corresponding studies of adiabatic expansions for fields of different spins and masses which necessitate higher order counterterms).

\section{Renormalized stress tensors (and $N_{\rm eff}$ bounds)}

We turn our attention in this section to the second and most consequential step in the process of renormalization -- that of extracting physical observables after imposing renormalization conditions. We begin the discussion with the relatively academic exercise of renormalizing the stress tensor of a test scalar field on a background that transitions in and out of inflation. By virtue of not having a classically evolving background and energy density, the only effects of a test scalar will be in renormalizing background couplings in the context of the effective theory of gravity \cite{Donoghue:1994dn, Donoghue:1995cz, Donoghue:2012zc, Burgess:2003jk}. We return to the more interesting and physically relevant case of primordial gravitational waves next. 

Our starting point is the bare matter and gravitational actions, along with the requisite counterterms:
\eqn{act0}{S = S_{\rm EH} + S_{\rm bg} + S_{\phi} + S_{\rm ct}.}
In order to impose renormalization conditions after having regularized divergences, we first consider the background equations of motion
\eqn{EOM1}{\frac{1}{8 \pi G_B}\left(R_{\mu \nu} -\frac{1}{2} R g_{\mu \nu} \right) = T^{\rm bg}_{\mu \nu} + T^{\phi}_{\mu \nu} + T^{\rm ct}_{\mu \nu},}
where we stress that the couplings that appear in Eq. \ref{act0} are to be understood as bare couplings, and the presence of the counterterm shifts the tadpole condition in a manner that we will shortly make precise. Given the consistency of dimensional regularization with general covariance, we can proceed by considering the above for any given component. For the 00 component, we have
\eqn{EOM2}{-\frac{R_{0}{}^{ 0}}{8 \pi G_B} = \rho^{\rm cl}_{\rm bg} + \rho_\phi + \rho_{\rm ct},}
where  $\rho^{\rm cl}_{\rm bg}$ is the to be renormalized classical background energy density that sources the expanding geometry around which we have computed the stress tensor for the test scalar field, with corresponding energy density $\rho_\phi$ defined as
\eqn{}{\rho_\phi := \rho^{\rm cl}_\phi - \langle T^0{}_0\rangle \equiv - \langle T^0{}_0\rangle,}
where $\rho_\phi^{\rm cl} \equiv 0$ for a test scalar by assumption. The background satisfies 
\eqn{}{-R_{0}{}^{ 0}=-\frac{3}{a^2} \left( \frac{a^{\prime}}{a}\right)^{\prime} = \frac{3H^2}{(a/a_{\rm R})^4}}
via Eq. \ref{sfdef} during the terminal radiation dominated phase, where it is to be stressed again that $H$ is the Hubble constant during the intermediate phase of inflation that enters the definition of the scale factor during radiation domination via Eq. \ref{sfdef}. We need to identify the counterterm that absorbs the divergence exhibited in Eq. \ref{rholog}: 
\eqn{EOM3.1}{\frac{1}{8 \pi G_B} \frac{3  H^2}{(a/a_R)^4} = \rho^{\rm cl}_{\rm bg} + \frac{H^4(1 + e^{-4\mathcal N_{\rm tot}})}{8\pi^2 (a/a_{\rm R})^4 } \left[  \frac{1}{\delta_{\rm UV}} + 1 - \gamma_E + \log\left(\frac{\mu}{H}\right) \right] + \rho_{\rm ct} + \rho_{\phi, \rm finite},}
where $\rho_{\rm finite} \equiv \rho_\phi - \rho_{\rm \phi, div}$. We are immediately presented with a choice here -- after subtracting the pole, do we proceed to (multiplicatively) renormalize Newton's constant, or (additively) renormalize the background whose unshifted value is determined by $\rho^{\rm cl}_{\rm bg}$? It turns out that this choice is rendered moot by the tadpole condition that determines the background equations of motion at any given order in $\hbar$, in that whichever choice we make will lead us to the same shifted tadpole condition. We thus proceed by reabsorbing the pole by adding a counterterm that multiplies the Ricci scalar in the Einstein Hilbert action with coefficient $B$ defined as 
\eqn{}{\rho_{\rm ct} = \frac{3  H^2}{(a/a_R)^4}\left(\frac{B_{-1}}{\delta_{\rm UV}} + B_0  \right).}
By assigning 
\eqn{count}{ B_{-1} = -\frac{H^2 (1 + e^{-4\mathcal N_{\rm tot}})}{24\pi^2 },}
one subtracts the pole contribution. By defining a scale dependent gravitational coupling as
\eqn{phys0}{\frac{1}{8\pi G_N(\mu)}=\frac{1}{8\pi G_B} - B_0 - \frac{H^2}{24\pi^2} (1 + e^{-4\mathcal N_{\rm tot}})\left\{1 - \gamma_E + \log\left(\frac{\mu}{H}\right) \right\},}
we can rewrite Eq. \ref{EOM3.1} as
\eqn{EOM3}{\frac{1}{8 \pi G_N(\mu)} \frac{3  H^2}{(a/a_R)^4} = \rho^{\rm cl}_{\rm bg} + \rho_{\rm \phi, finite},}
where the finite remainder from the counterterm and $\rho_{\phi, \rm div}$ are absorbed by the scale dependent gravitational coupling, allowing us to start imposing renormalization conditions to fix the finite parts. We do so by determining the Newtonian constant via a measurement at some energy scale $\mu_*$, with which we can eliminate all reference to $G_B$ and $B_0$ via
\eqn{phys2}{\frac{1}{8\pi G_B}=\frac{1}{8\pi G_N(\mu_*)} + B_0 + \frac{H^2}{24\pi^2} (1 + e^{-4\mathcal N_{\rm tot}})\left\{1 - \gamma_E + \log\left(\frac{\mu_*}{H}\right) \right\},}
and substituting the result into Eq. \ref{phys0} to obtain 
\eqn{phys3}{\frac{1}{8\pi G_N(\mu)}=\frac{1}{8\pi G_N(\mu_*)}- \frac{H^2}{24\pi^2 }(1 + e^{-4\mathcal N_{\rm tot}}) \log\left(\frac{\mu}{\mu_*}\right).}
Picking $\mu_*$ to be some scale where we have determined  Newton's constant\footnote{Note that measuring the strength of the gravitational coupling can only be done via a Cavendish type experiment, typically done at laboratory scales where we have independent knowledge of the masses whose mutual gravitational force we can determine. This yet another manner in which gravity is distinguished among forces as the only force whose coupling strength we measure in the UV (i.e. mm scale) and run into the IR, rather than the other way around.} to be $8\pi G_{N}(\mu_*) = \Mpl^{-2}$ where $\Mpl = 2.435 \times 10^{18}$ GeV, we can finally express $G_{N}(\mu)$ as
\eqn{}{8\pi G_N(\mu) = \frac{1}{\Mpl^2}\left[1 - \frac{H^2(1 + e^{-4\mathcal N_{\rm tot}})}{24\pi^2\Mpl^2 } \log\left(\frac{\mu}{\mu_*}\right)\right]^{-1},}
which can be used to express Eq. \ref{EOM3} in its fully covariant form as
\eqn{}{G_{\mu\nu} = \frac{1}{\Mpl^2}T^{\rm bg, shift}_{\mu\nu}\left[1 - \frac{H^2(1 + e^{-4\mathcal N_{\rm tot}})}{24\pi^2\Mpl^2 } \log\left(\frac{\mu}{\mu_*}\right)\right]^{-1},}
where the shifted background stress tensor is defined as the sum of tree level and finite contributions on the left hand side of Eq. \ref{EOM3}. Several things are to be immediately noted here -- foremost is the minuscule nature of the scale dependence of the gravitational coupling should we phrase it that way. We could also simply view it as a multiplicative renormalization of the background matter content that sources the expansion history\footnote{This interpretation is to be preferred if one would like keep the graviton to be canonically normalized throughout all of cosmic history, something that is implicitly taken for granted for most quoted observational results.}. Given that virtual effects from test scalar fields serve only to renormalize background quantities\footnote{Noting that given the vanishing background energy density of the test scalar field, it contributes vanishingly to the curvature perturbation.} and impart scale dependence in observables associated to other propagating degrees of freedom (and that too, in a highly suppressed manner \cite{delRio:2018vrj, Baumgart:2021ptt}), this is the furthest we can take this exercise. The situation for gravitational waves is more interesting. 

Retracing the steps above with the dimensionally regularized result for gravitational waves on a finite duration background in Eq. \ref{drimp}, we end up with the renormalized background 
\eqn{finalren}{G_{\mu\nu} = \frac{1}{\Mpl^2}T^{\rm bg, shift}_{\mu\nu}\left[1 - \frac{H^2(1 + e^{-4\mathcal N_{\rm tot}})}{12\pi^2\Mpl^2 } \log\left(\frac{\mu}{\mu_*}\right)\right]^{-1},}
which should be the starting point for determining any constraints on vacuum sourced primordial gravitational waves from $N_{\rm eff}$ bounds. The latter is in essence the question of how vacuum tensor perturbations renormalize the background expansion through $1/a^4$ contributions that mimic additional relativistic species. Unlike the case for virtual test scalars, however, gravitational waves have a classically evolving background upon which they represent perturbations -- the background geometry itself. Therefore, we have additional means to potentially measure the contributions from vacuum tensor modes. 

We first reconsider the spectral density for gravitational waves $\Omega^{\rm gw}(k,\tau)$, whose amplitude on the scale invariant plateau for sub-horizon modes is well defined, and given by 
\eqn{}{\Omega^{\rm gw}(k,\tau) = \frac{H^4}{16\pi^4 (a/a_{\rm R})^4}~~~~~ {\rm (sub-horizon)}}
as plotted in Fig. \ref{fig:gwIsc}. Let us presume that we have the means to determine the ratio $H^2/\Mpl^2$ during inflation via measurement of the tensor to scalar ratio at some pivot scale via B-mode anisotropy observations, or via the measurement of the spectral density of the stochastic gravitational wave background at some fixed scale via interferometeric means, or both\footnote{In practice, both B-mode polarization measurements and the detection of stochastic backgrounds through interferometry measure momentum transfer of gravitational waves and not the energy density associated with them. Consequently the relevant quantity one can infer by such measurements would be the analog of the Poynting vector associated with a flux of gravitational waves, which would nevertheless still be sensitive to the ratio $H^2/\Mpl^2$.}. In the context of Eq. \ref{finalren}, which can now be re-expressed as 
\begin{eqnarray} \nonumber
\frac{3  H^2}{(a/a_R)^4} &=& \frac{1}{\Mpl^2}\left(\rho_{\rm bg}^{\rm cl} + \rho_{\rm gw, finite}\right)\left[1 - \frac{H^2(1 + e^{-4\mathcal N_{\rm tot}})}{12\pi^2\Mpl^2 } \log\left(\frac{\mu}{\mu_*}\right)\right]^{-1}\\ \label{GWfin}
&=& \frac{\rho_{\rm bg}^{\rm cl}}{\Mpl^2}\left(1 + \delta_{\rm r}\right)\left[1 - \frac{H^2(1 + e^{-4\mathcal N_{\rm tot}})}{12\pi^2\Mpl^2 } \log\left(\frac{\mu}{\mu_*}\right)\right]^{-1},  
\end{eqnarray}
where $\delta_{\rm r}$ is a constant since the putative tree level background and the stochastic background of vacuum tensor modes both scale as $\propto (a/a_{\rm R})^4$. Therefore, recalling that the quantities that parameterized the background on the left hand side of Eq. \ref{EOM2} were also implictly bare quantities, one finds that the shifted tadpole condition one obtains upon renormalization is indistinguishable from a rescaling of the scale factor normalization at reheating, or to simple shift the temperature redshift relation in an otherwise unobservable manner\footnote{We should also stress that although we kept the scale dependent factor in Eq. \ref{GWfin} for completeness, it can effectively be set to unity given the current upper bounds on the tensor to scalar ratio $r_*$ or about $r_* \lesssim 3 \times 10^{-2}$ \cite{Campeti:2022vom, BICEP:2021xfz, BICEP2:2018kqh}, so that $H^2/(8\pi^2\Mpl^2) = \frac{r_*}{16}\Delta^2_{\mathcal R}\lesssim 10^{-12},$ in combination with the fact that the log of the ratio between laboratory and Hubble scales is no more than order $10^2$.}. 

Therefore, vacuum tensor fluctuations by their very nature only serve to renormalize background quantities, and do not enter as an additional effective light species as registered by $N_{\rm eff}$ bounds. This should be immediately apparent from the physical nature of $N_{\rm eff}$ bounds as measuring the ratio of propagating light species that have undergone freeze-out relative to the entropy density of the universe, which does not apply to vacuum fluctuations. This is of course, not true for gravitons that are physically produced by some mechanism in the early universe, however the latter will also feature a bounded integrated spectral density.   

\section{Concluding remarks}
Divergences in primordial observables are not something cosmologists have the luxury of ignoring: every cosmological tracer corresponding to a density fluctuation samples and convolves the coincident limit of a field bilinear, necessitating subtraction. Merely regularizing, however, is not enough. The process of arriving at a physical observable is incomplete unless one follows through by constructing the requisite counterterms and fixing any finite contributions that could accompany any subtraction via the imposition of renormalization conditions. Failure to do so runs the risk of drawing unphysical conclusions that include scheme (e.g. cutoff) dependence in observables where there should not be any, or over-interpreting contributions to physical observables that are absorbed or otherwise accounted for in the process of renormalization. 

Fortunately, the regularization and renormalization of divergences on cosmological backgrounds is a well understood if often elided process. In following through the details of this procedure at some length, we hope to have guided the reader through how this works in practice in a foliation specific formalism that should be familiar to most cosmologists, uncovering various novelties along the way. Not only have we elaborated on how to make sense of what one might obtain from putting hard cutoffs in physical momenta relative to what one might have obtained from dimensional regularization, we provided demonstrations of the scheme independence of logarithmic divergences and explicitly tracked this process on backgrounds that have a beginning and end to inflation. Certain nominally IR divergent quantities on backgrounds that model inflation as past infinite dS space are cured when one considers finite duration inflation. Whether this generalizes beyond the simple examples elaborated upon this review to interacting theories seems like a highly pertinent question to follow up on. We furthermore stressed the need to work with an improved stress tensor for gravitational waves that does not presume a prior scale separation in the context of cosmology, and demonstrated how the process of attempting to extract $N_{\rm eff}$ bounds from vacuum tensor fluctuations is inextricable from the process of background renormalization. 

\bmhead{Acknowledgements}

We are grateful to Paolo Benincasa and Koenraad Schalm for many discussions and comments on the draft. We are especially indebted to Cliff Burgess for valuable discussions and  insights that permeate this review, and to Alessandro Bettini for his patience with the manuscript. AN is grateful to Dimitrios Krommydas and Alice Garoffolo for the useful discussions, and the Casimir Research School for travel funds to visit the Perimeter Institute, where some of the work leading to this review was initiated.

\begin{appendices}

\section{\label{app:emt} Dimensional regularization of stress tensors in FRLW}
In this appendix, we show how dimensional regularization renders covariant counterterms for the divergences associated with the stress energy tensors for scalar fields and tensor perturbations in a foliation specific formalism. We first consider the example of a massive scalar field on an FRLW background, and consider the analogs of Eqs. \ref{rmfds} and \ref{pmfds} in cosmic time where we furthermore neglect the effects of background expansion. This is in order to be facilitate working with transparent analytic expressions that moreover, are non-vanishing (as would be the case for a massless scalar field), with the further justification that as we are only interested in computing the relevant counterterms to subtract UV divergences and can thus be forgiven for this approximation for illustrative purposes. Details of how one dimensional regularizes energy momentum tensors for more realistic examples in a fully covariant approach can be found in e.g. \cite{Birrell:1982ix, Parker:2009uva}. What follows below closely tracks the treatment of \cite{Collins:1984xc, Koksma:2011cq}.

Working in $D= 4 - \delta$ dimensions, where the background metric is of the FRLW form in cosmic time, one finds the following expressions for the vacuum expectation values of the stress tensor:
\begin{eqnarray} \nonumber
\left\langle 0\left|\hat{T}_{00} \right| 0\right\rangle & =& \frac{\mu^{4-D}}{2 a^{D-1} (2 \pi)^{D-1}} \int \mathrm{d}^{D-1} k \sqrt{m^2+\frac{k^2}{a^2}} \\ \label{stcomp}
		\left\langle 0\left|\hat{T}_{i i}\right| 0\right\rangle & =& \frac{\mu^{4-D}}{2 a^{D-1} (2 \pi)^{D-1}} \frac{1}{D-1} \int \mathrm{d}^{D-1} k \frac{k^2}{\sqrt{m^2+\frac{k^2}{a^2}}},
\end{eqnarray}
both components of which nominally have the same degree of divergence in the UV, but with different coefficients. Imposing hard cutoffs in physical momenta would necessitate a counterterm that cannot be constructed from background geometric invariants. Instead, we proceed via Eq. \ref{Ians}, with $A=0$, $B=-\frac{1}{2}$ in the first integral above, and $A=1$, $B=\frac{1}{2}$ for the second integral, to find
\begin{eqnarray} \nonumber
		&& \int \mathrm{d}^{D-1} k \sqrt{m^2+\frac{k^2}{a^2 }} = \frac{\left(m a\right)^{D} }{a} \frac{(2 \pi)^{D-1}}{(4 \pi)^{\frac{D-1}{2}}} \frac{\Gamma \left(\frac{D-1}{2}\right) \Gamma \left(-\frac{D}{2}\right)}{\Gamma \left(-\frac{1}{2}\right) \Gamma \left(\frac{D-1}{2}\right)} \\
		&& \int \mathrm{d}^{D-1} k \frac{k^2}{\sqrt{m^2+\frac{k^2}{a^2 }}} = a \left(m a\right)^{D}\frac{(2 \pi)^{D-1}}{(4 \pi)^{\frac{D-1}{2}}} \frac{\Gamma \left(\frac{1}{2} + \frac{D}{2}\right) \Gamma \left(-\frac{D}{2}\right)}{\Gamma \left(\frac{1}{2}\right) \Gamma \left(\frac{D-1}{2}\right)}, 
\end{eqnarray}
so that the dimensionally regularized components of the stress energy tensor become
\begin{eqnarray} \nonumber
\left\langle 0\left|\hat{T}_{00} \right| 0\right\rangle && =\frac{\mu^{4-D}}{2 a^{D-1} (2 \pi)^{D-1}} \frac{\left(m a\right)^{D} }{a} \frac{(2 \pi)^{D-1}}{(4 \pi)^{\frac{D-1}{2}}} \frac{\Gamma \left(\frac{D-1}{2}\right) \Gamma \left(-\frac{D}{2}\right)}{\Gamma \left(-\frac{1}{2}\right) \Gamma \left(\frac{D-1}{2}\right)} \\ \nonumber
		&& =\frac{\mu^{4-D}}{2 (4 \pi)^{\frac{D-1}{2}} } \frac{\left(m a\right)^{D} }{ a^{D}}\frac{\Gamma \left(-\frac{D}{2}\right)}{\Gamma \left(-\frac{1}{2}\right)} \\ 
		&& =\frac{\mu^{4}}{2 (4 \pi)^{\frac{D-1}{2}} } \left(\frac{m  }{ \mu} \right)^{D}\frac{\Gamma \left(-\frac{D}{2}\right)}{\Gamma \left(-\frac{1}{2}\right)},
\end{eqnarray}
for the energy density, and 
\begin{eqnarray} \nonumber
		\left\langle 0\left|\hat{T}_{i i}\right| 0\right\rangle && =\frac{\mu^{4-D}}{2 a^{D-1} (2 \pi)^{D-1}} \frac{1}{D-1}a \left(m a\right)^{D} \frac{(2 \pi)^{D-1}}{(4 \pi)^{\frac{D-1}{2}}} \frac{\Gamma \left(\frac{1}{2} + \frac{D}{2}\right) \Gamma \left(-\frac{D}{2}\right)}{\Gamma \left(\frac{1}{2}\right) \Gamma \left(\frac{D-1}{2}\right)} \\ \nonumber
		&& =\frac{\mu^{4-D}}{2 (4 \pi)^{\frac{D-1}{2}} } \frac{1}{D-1} \frac{a \left(m a\right)^{D}}{a^{D-1} } \frac{\Gamma \left(\frac{D+1}{2}\right) \Gamma \left(-\frac{D}{2}\right)}{\Gamma \left(\frac{1}{2}\right) \Gamma \left(\frac{D-1}{2}\right)} \\
		&& = -\frac{\mu^{4} a^2 }{2 (4 \pi)^{\frac{D-1}{2}} } \left(\frac{m  }{ \mu} \right)^{D}\frac{ \Gamma \left(-\frac{D}{2}\right)}{\Gamma \left(-\frac{1}{2}\right) }, 
\end{eqnarray}
for the pressure components, where in the last equality we have used $\Gamma (x+1) = x \Gamma (x) $ to re-express 
\eqn{}{\frac{\Gamma \left(\frac{D+1}{2}\right)}{\Gamma \left(\frac{1}{2}\right) \Gamma \left(\frac{D-1}{2}\right)} = -\frac{D-1}{\Gamma \left(-\frac{1}{2}\right)}.}
From this, we see that the stress tensor has a divergence of the form
\eqn{}{\langle \hat{T}_{\mu\nu} \rangle_{\rm div} = - g_{\mu\nu} \frac{m^4}{64\pi^2}\left\{\frac{2}{\delta} - \log\left(\frac{m^2}{4\pi\mu^2}\right) + \gamma_E - \frac{3}{2} \right\},}
which can straightforwardly be subtracted by a cosmological constant-like counterterm.

Similarly, we follow through the same logic to study the divergencies appearing in computing the energy density and pressure of gravitational waves. We find that even if at the operator level this is not evident, once one regularize using dimensional regualrization the equation of state ($P=\frac{1}{3} \rho$ in the case under analysis) is satisfied and the trace of the energy tensor is proportional to the Ricci scalar ($T_{\mu}{}^{\mu}=R=0$ in the case under analysis), consistent with subtraction with a counterterm that comes from varying the background curvature with respect to the metric (and thus corresponds to a renormalization of $G_{ N}$, cf. the discussion following Eq. \ref{drimp0}).
 
Following the procedure decribed in Appendix D, we derive the energy density and pressure of gravitational waves during RD era that result respectivly
\begin{eqnarray} \nonumber
		\rho_{\rm gw}=&&\frac{1}{8 \pi a^2 G_N}\left\langle \frac{1}{8} h^\prime_{j}{}^{ i} h^\prime_{i}{}^{  j}- \frac{3}{8} \partial_k h_{ij} \partial^k h^{i j} - \frac{1}{2}  h_{ij} \partial_k\partial^k h^{i j}+  \mathcal{H}h_{j}{}^{ i} h^\prime_{i}{}^{ j}+\frac{1}{4} \partial_k h_{ij} \partial^j h^{ ik} \right\rangle   \\\nonumber
		P_{\rm gw}=&&-\frac{1}{3}  \frac{1}{8 \pi a^2  G_N}\left\langle \frac{5}{8}h^\prime_{j}{}^{ i} h^\prime_{i}{}^{  j}- \frac{3}{8} \partial_k h_{ij} \partial^k h^{i j} - \frac{1}{2}  h_{i j}\partial_k\partial^k h^{ ij}+2  \mathcal{H}h_{j}{}^{ i} h^\prime_{i}{}^{ j} + h_{j}{}^{ i} h^{\prime \prime}_{i}{}^{  j}\right\rangle \\
		=&&\frac{1}{3} \frac{1}{8 \pi a^2  G_N}\left\langle - \frac{5}{8}h^\prime_{j}{}^{ i} h^\prime_{i}{}^{  j}+ \frac{3}{8} \partial_k h_{ij} \partial^k h^{i j} - \frac{1}{2}  h_{ij} \partial_k\partial^k h^{i j} \right\rangle 
\end{eqnarray}
where in the second equality of the pressure we use the EOM. From the result above it appears that the equation of state is not satisfied. However, specifying the result in the case of finite inflation and following the procedure in Section \ref{subsec:gwfi}, we find that by using physical cutoff and dimensional regularization respectively,  the regularized pressure results 
\vspace{-0.2cm}
\eqn{}{ P_{\rm gw, div} = \frac{1}{3}  \lim_{\Lambda_{\rm UV} \to \infty}  \left\{ \frac{1}{2  \pi^2 } \frac{\Lambda_{\rm UV}^4}{4}  -\frac{5}{2} \frac{ a_{\rm R}^4 H^2}{2  \pi^2 a^4} \frac{\Lambda_{\rm UV}^2}{2} +\frac{  H^4\left(1 + e^{-4\mathcal N_{\rm tot}}\right) }{4\pi^2 (a/a_{\rm R})^4}\log \frac{\Lambda_{\rm UV}}{\Lambda_{\rm IR}} \right\} ~ {\rm (cutoff)}}
\eqn{drimp}{ P_{\rm gw, div} = \frac{1}{3} \lim_{\delta_{\rm UV} \to 0}  \left\{ \frac{H^4\left(1 + e^{-4\mathcal N_{\rm tot}}\right)}{4\pi^2 (a/a_R)^4}  \left[  \frac{1}{\delta_{\rm UV}} + 1 - \gamma_E + \log\left(\frac{\mu}{H}\right) \right] \right\}. ~~~~~~~~~~~~ {\rm (dim-reg)} }
By recalling the result found in Section \ref{subsec:gwfi} for the energy density
\eqn{}{ \rho^{\rm Imp}_{\rm gw, div} =  \lim_{\Lambda_{\rm UV} \to \infty}  \left\{ \frac{1}{2  \pi^2 } \frac{\Lambda_{\rm UV}^4}{4}  -\frac{7}{2} \frac{ a_{\rm R}^4 H^2}{2  \pi^2 a^4} \frac{\Lambda_{\rm UV}^2}{2} +\frac{  H^4\left(1 + e^{-4\mathcal N_{\rm tot}}\right) }{4\pi^2 (a/a_{\rm R})^4}\log \frac{\Lambda_{\rm UV}}{\Lambda_{\rm IR}} \right\} ~~~~ {\rm (cutoff)} }
\eqn{}{ \rho^{\rm Imp}_{\rm gw, div} =  \lim_{\delta_{\rm UV} \to 0}  \left\{ \frac{H^4\left(1 + e^{-4\mathcal N_{\rm tot}}\right)}{4\pi^2 (a/a_R)^4}  \left[  \frac{1}{\delta_{\rm UV}} + 1 - \gamma_E + \log\left(\frac{\mu}{H}\right) \right] \right\}, ~~~~~~~~~~~~ {\rm (dim-reg)}}
we can see that only the result regularized using dimensional regularization gives a traceless stress energy tensor and satisfies the equation of state for radiation-like species.
\vspace{-0.2cm}
\section{Oscillatory contributions to divergent integrals \label{app:osc}}
In deriving various expressions in Section \ref{sec:fd} and beyond, we neglected oscillatory contributions coming from $\alpha^{\rm R*}_k\beta^{\rm R}_k $ as well as in computing the modulus square of $|\beta_k^{\rm R}|^2$ in final expressions for the UV divergences we are left to regulate, such as Eqs. \ref{2pt} and \ref{r2pt7}. In this appendix we show that this is justified focusing on the details for the scalar case, with the extension to analogous computations for gravitational waves a straightforward extension. We reconsider Eq. \ref{r2pt7} but with the neglected oscillatory contributions included:
\eqn{r2pt9}{ \rho = \frac{1}{8\pi^4 a^4}\int_{-\infty}^\infty \,\frac{d^4k}{k} \left[k + \frac{a_{\rm R}^4 H^2}{a^2 2 k} \right]\left\{1 +\frac{a_{\rm R}^4 H^4}{2k^4}+2   \left|  \beta_k^{\rm I} \right|^2\left(1+ \frac{a_{\rm R}^4 H^4}{2k^4}\right)\right\} + \rho_{\rm osc},}
where $\rho_{\rm osc}$ is given by
\begin{eqnarray}
	\nonumber \rho_{\rm osc} &=& \frac{1}{4\pi^2 a^4}\int_{0}^\infty \,dk \: k^2 \left(k + \frac{a_{\rm R}^4 H^2}{a^2 2 k} \right)\left[- i \alpha_k^{\rm I}  \beta_k^{\rm I*} e^{2i\frac{k}{a_{\rm R} H}}  \frac{a_{\rm R}^2 H^2}{k^2}  \left(-i + \frac{a_{\rm R} H}{k}+i \frac{a_{\rm R}^2 H^2}{2k^2}\right)
	\right.\\ 	\nonumber
	&&\left. +i  \alpha_k^{\rm I*}  \beta_k^{\rm I} e^{-2i\frac{k}{a_{\rm R} H}}  \frac{a_{\rm R}^2 H^2}{k^2} \left(i + \frac{a_{\rm R} H}{k}-i \frac{a_{\rm R}^2 H^2}{2k^2}\right) \right] \\
	&&+\nonumber \frac{1}{4\pi^2 a^4}\int_{0}^\infty \,dk \: k^2  \left[\alpha^{\rm R}_k\beta^{\rm R*}_k e^{-2 i k\tau_{\rm R} \left( 2-\frac{a}{a_{\rm R}} \right)} \left(\frac{k}{2} +\frac{a_{\rm R}^4 H^2}{a^2 2 k} \left(1 + i\frac{k a}{a_{\rm R}^2 H}\right)^2\right) \right. \\ 	\label{roscpt1}
	&&+\left. \alpha^{\rm R*}_k\beta^{\rm R} e^{2 i k\tau_{\rm R} \left( 2-\frac{a}{a_{\rm R}} \right)} \left(\frac{k}{2} +\frac{a_{\rm R}^4 H^2}{a^2 2 k} \left(1 -i\frac{k a}{a_{\rm R}^2 H}\right)^2\right)   \right] .   
\end{eqnarray}
Eq. \ref{r2pt7} presumes that $\rho_{\rm osc}$ renders vanishing contributions. Making use of Eqs. \ref{mcI2} and \ref{mcR}, one can integrate the various terms and expand the result in the limit $k \to \infty$, upon which we obtain 
\vspace{-0.2cm}
\eqn{}{\begin{aligned}
	\rho_{\rm osc} &=\frac{1}{4\pi^2 a^4} \lim_{k \to \infty} \left[ \left(e^{ \left(-\frac{2 i k (a_{\rm I}-a_{\rm R}) }{H a_{\rm R}^2 }\right)}+e^{ \left(\frac{2 i k (a_{\rm I}-a_{\rm R}) }{H a_{\rm R}^2 }\right)}\right)  \left( -\frac{H^4 a_{\rm R}^6}{4 a^2-4 a a_{\rm R}} \right) \right. \\ 	\label{roscpt2}
 &\left.+ \left(e^{ \left(-\frac{2 i k (a a_{\rm I}-2 a_{\rm I} a_{\rm R}+ a_{\rm R}^2) }{H a_{\rm I} a_{\rm R}^2 }\right)}+e^{ \left(\frac{2 i k (a a_{\rm I}2 a_{\rm I} a_{\rm R}+ a_{\rm R}^2) }{H a_{\rm I} a_{\rm R}^2 }\right)}\right)  \left( \frac{H^4 a_{\rm I}^3 a_{\rm R}^4}{4 a\left(a a_{\rm I}-2 a_{\rm I} a_{\rm R}+a_{\rm R}^2\right)}\right) +... \right].
\end{aligned}}
up to terms of order $\sim \frac{1}{k} $. We note that this contribution is manifestly finite, and oscillatory in the upper limit in a manner that would be eliminated via the analog of the $i\epsilon$ prescription for the Minkowski space limit of all the two point propagators in the in-in formalism. Similarly, for the tensor perturbations, the full expression for the energy density including the oscillatory contributions is given for the Isaacson form of the stress tensor Eq. \ref{Rho gw literature 0}, by
\vspace{-0.2cm}
\begin{eqnarray}
	\nonumber
	\rho^{\rm Isc}_{\rm gw} &= &\frac{1}{4\pi^4 a^4}   \int_{-\infty}^\infty \,\frac{d^4k}{k} \left[k + \frac{ a_{\rm R}^4 H^2}{a^2  k} \right]\left\{1 +\frac{a_{\rm R}^4 H^4}{2k^4}+2   \left|  \beta_k^{\rm I} \right|^2\left(1+ \frac{a_{\rm R}^4 H^4}{2k^4}\right)
	~~~({\rm Isaacson})\right.\\ 	\nonumber
	&& \left. - i \alpha_k^{\rm I}  \beta_k^{\rm I*} e^{2i\frac{k}{a_{\rm R} H}}  \frac{a_{\rm R}^2 H^2}{k^2}  \left(-i + \frac{a_{\rm R} H}{k}+i \frac{a_{\rm R}^2 H^2}{2k^2}\right) +i  \alpha_k^{\rm I*}  \beta_k^{\rm I} e^{-2i\frac{k}{a_{\rm R} H}}  \frac{a_{\rm R}^2 H^2}{k^2} \right.\\ 	\nonumber
	&& \left. \left(i + \frac{a_{\rm R} H}{k}-i \frac{a_{\rm R}^2 H^2}{2k^2}\right) \right\} +\nonumber \frac{1}{4 \pi^4 a^4}  \int_{-\infty}^\infty \,\frac{d^4k}{k}  \left(\frac{a_{\rm R}^4 H^2}{a^2  k} \right)  \left[\alpha^{\rm R}_k\beta^{\rm R*}_k e^{-2 i k\tau_{\rm R} \left( 2-\frac{a}{a_{\rm R}} \right)} \right.\\ 	  \label{rhogwfull} 
	&& \left.  \left(1+i\frac{k a}{a_{\rm R}^2 H}\right)^2 + \alpha^{\rm R*}_k\beta^{\rm R} e^{2 i k\tau_{\rm R} \left( 2-\frac{a}{a_{\rm R}} \right)}  \left(1-i\frac{k a}{a_{\rm R}^2 H}\right)^2  \right],   
\end{eqnarray}
and for the improved form of the stress tensor Eq. \ref{eq:rho 1 main}, as:
\vspace{-0.2cm}
\begin{eqnarray}
		\nonumber
	\rho_{\rm gw}^{\rm Imp} &=& \frac{1}{4\pi^4{a^4}} \int_{-\infty}^\infty \,\frac{d^4k}{k} \left[k - \frac{7 a_{\rm R}^4 H^2}{a^2 2 k} \right]\left\{1 +\frac{a_{\rm R}^4 H^4}{2k^4}+2   \left|  \beta_k^{\rm I} \right|^2\left(1+ \frac{a_{\rm R}^4 H^4}{2k^4}\right)
	\right.~({\rm Improved})\\ 
	&&\nonumber \left. - i \alpha_k^{\rm I}  \beta_k^{\rm I*} e^{2i\frac{k}{a_{\rm R} H}}  \frac{a_{\rm R}^2 H^2}{k^2}  \left(-i + \frac{a_{\rm R} H}{k}+i \frac{a_{\rm R}^2 H^2}{2k^2}\right)+i  \alpha_k^{\rm I*}  \beta_k^{\rm I} e^{-2i\frac{k}{a_{\rm R} H}}  \frac{a_{\rm R}^2 H^2}{k^2} \right.\\ 	\nonumber
	&& \left.  \left(i + \frac{a_{\rm R} H}{k}-i \frac{a_{\rm R}^2 H^2}{2k^2}\right) \right\} + \frac{1}{4\pi^4{a^4}}  \int_{-\infty}^\infty \,\frac{d^4k}{k}  \left[\alpha^{\rm R}_k\beta^{\rm R*}_k e^{-2 i k\tau_{\rm R}  \left( 2-\frac{a}{a_{\rm R}} \right)}   \right.\\ 	\nonumber && \left. \left(\frac{k}{2} +\frac{a_{\rm R}^4 H^2}{a^2 2 k} \left(1+i\frac{k a}{a_{\rm R}^2 H}\right)^2 -4 \frac{a_{\rm R}^4 H^2}{a^2  k} \left(1+i\frac{k a}{a_{\rm R}^2 H}\right)  \right) + \alpha^{\rm R*}_k\beta^{\rm R} e^{2 i k\tau_{\rm R} \left( 2-\frac{a}{a_{\rm R}} \right)}  \right.\\ \label{rhoimproved}
	&& \left.  \left(\frac{k}{2} +\frac{a_{\rm R}^4 H^2}{a^2 2 k} \left(1-i\frac{k a}{a_{\rm R}^2 H}\right)^2-4 \frac{a_{\rm R}^4 H^2}{a^2  k} \left(1-i\frac{k a}{a_{\rm R}^2 H}\right) \right)  \right] ,
\end{eqnarray}
which again results in oscillating but finite contributions in the UV-limit that can similarly be discarded. Importantly, even though the oscillatory contributions are negligible in the UV, they become relevant when the oscillations freeze as $k\to 0$, also softening the IR behavior for the spectral power density relative to what would have been on a past infinite dS background.  
\vspace{-0.2cm}
\section{\label{app:ps} Point split comparison to physical cutoff regularization}

In what follows we compare the results of previous sections had we commuted taking the coincidence limit before integrating over $\vec{k}$ to show agreement between point split regularization and physical momentum cutoff regularization. Recalling the steps that led up to Eq. \ref{pscon}, we are interested in computing quantities that derive from the coincidence limit of the two point function
\vspace{-0.2cm}
\begin{eqnarray}\nonumber
	\lim_{x \to y} G(x,y)&=& \lim_{x \to y} \langle \hat{\phi}(\tau,x) \hat{\phi} (\tau',y) \rangle \\ \label{2pt2}
	 &=& \lim_{x \to y}  \int \frac{d^3k}{(2\pi)^3} \: \phi(\tau, k)  \phi^*(\tau, k)  e^{i{k}\cdot ({x}-{y})},
\end{eqnarray}
where in the second equality we use the Fourier transform convention of Eq. \ref{fourier}. Similarly, in the derivation of the energy density for a massless test scalar field, we are interested in computing the coincidence limit
\vspace{-0.4cm}
\eqn{rholim}{\begin{aligned}
	\rho(x)=\lim_{x \to y} \rho(x,y)&=\lim_{x \to y} \frac{1}{2a^2}\left[ \langle \hat{\phi}^{\prime}(\tau,x)\hat{\phi}^{\prime}(\tau',y) \rangle+ \langle {\nabla}_x \hat{\phi}(\tau,x){\nabla}_y \hat{\phi}(\tau',y) \rangle \right]\\
	 &=\lim_{{x} \to {y}}  \frac{1}{2a^2}\int \frac{d^3k}{(2\pi)^3} \: \left[ \phi^{\prime}(\tau, k)  \phi^{*\prime}(\tau, k) + k^2 \phi(\tau, k)  \phi^*(\tau, k)  \right] e^{i{k}\cdot ({x}-{y})}.
\end{aligned}}
Defining $r := |x-y|$, we now look to evaluate the following point split quantities
\vspace{-0.2cm}
\begin{eqnarray}
	\nonumber
	&& G(x,x)=\lim_{r \to 0} \int \frac{d^3k}{(2\pi)^3} \: \phi(\tau, k)  \phi^*(\tau, k)  e^{i{k}\cdot {r}}\\ \label{psquant}
	  &&\rho(x)=\lim_{r \to 0}  \frac{1}{2a^2} \int \frac{d^3k}{(2\pi)^3} \: \left[ \phi^{\prime}(\tau, k)  \phi^{*\prime}(\tau, k) + k^2 \phi(\tau, k)  \phi^*(\tau, k)  \right] e^{i{k}\cdot {r}}.
\end{eqnarray}
We note that in the derivations in the main body of this article, we freely commuted the limit $\lim_{r \to 0}$ with the integral over $\vec{k}$. However, this is strictly speaking only a well defined procedure if the integral is absolutely convergent. In what follows, we compute the integral over $\vec{k}$ before taking the coincidence limit. We start by computing Eq. \ref{2pt2} for a non-interacting, minimally coupled massless scalar field on a dS background, obtaining 
\vspace{-0.4cm}
\begin{eqnarray} \nonumber
	G(x,x)&=&\lim_{r \to 0}\frac{1}{(2\pi)^3} \int_0^{\infty} dk k^2 \int_{-1}^1 d \cos \theta \int_0^{2 \pi} d\varphi \: \frac{H^2}{2 k^3} \left( 1 + k ^2 \tau^2 \right) e^{i k r \cos \theta } \\
	\nonumber &=& \lim_{r \to 0} \frac{H^2}{4\pi^2} \frac{1}{r} \int_0^{\infty} dk \: \left( \frac{1}{k^2}  +  \tau^2 \right) \sin kr\\
	\nonumber &=& \lim_{r \to 0}  \frac{H^2}{4\pi^2} \left( \frac{\tau^2}{r^2}  - \left[\gamma_E - 1 + \log k r  \right]\right)\\ \label{2pt3ps}
	 &=& \lim_{\sigma \to 0}  \frac{H^2}{4\pi^2} \left( \frac{1}{H^2 \sigma^2}  - \left[\gamma_E - 1 + \log k_p \sigma  \right]\right),
\end{eqnarray}
where the third equality involves use of the principle value expression $\int_0^{\infty} dx \sin x =1$, and the last equality is expressed in terms of the physical distance $\sigma = a r$, with $k_p:= k/a$. As expected, Eq. \ref{2pt3ps} shows that the two point function diverges in the coincident limit $\sigma \to 0$ in the same manner as Eq. \ref{dsdiv00} obtained by imposing hard momentum cutoffs. In a similar manner, we can repeat the computation for the energy density given by Eq. \ref{rholim}, obtaining 
\begin{eqnarray} \nonumber
	\rho_{\rm div}&=&\lim_{r \to 0}\frac{1}{(2\pi)^3} \frac{H^2}{4 a^2} \int_0^{\infty} \frac{dk}{k} k^2 \int_{-1}^1 d \cos \theta \int_0^{2 \pi} d\varphi \:  \left( 1 + 2 k ^2 \tau^2 \right) e^{i k r \cos \theta } \\
	\nonumber &=& \lim_{r \to 0} \frac{1}{4 \pi^2} \frac{H^2}{2 a^2} \frac{1}{r}  \int_0^{\infty} dk \: \left( 1  + 2 k^2 \tau^2 \right) \sin kr\\
	\nonumber &=& \lim_{r \to 0} \frac{H^2}{2 a^2}  \frac{1}{4 \pi^2 r^2} \left[1 - \frac{4\tau^2}{r^2} \right]\\
	\label{rholimDS} &=& \lim_{\sigma \to 0} \frac{H^2}{8\pi^2}  \frac{1}{\sigma^2} \left[1 - \frac{4}{H^2\sigma^2} \right],
\end{eqnarray}
where going from the second to the third line involves integrating by parts and again using the principal value scheme. Again, we reproduce the divergences of Eq. \ref{rdshc} obtained by hard cutoffs in physical momenta, up to the expected scheme dependence which as we will show shortly, which drops out when computing the coefficients of logarithmic divergences. The comparison we are most interested in exhibiting this, however, is for the the case of finite duration of inflation. We  exhibit the procedure for the energy density given that the treatment for the two point function proceeds identically. Using the results of Eqs. \ref{modRD2} and \ref{r2pt7} in the context of Eq. \ref{psquant}, we find  
\begin{eqnarray} \nonumber
	 \rho_{\rm div} &=& \lim_{r \to 0} \frac{1}{4 \pi^2 a^4} \frac{1}{r}  \int_0^{\infty} dk  k^2 \left(1 + \frac{a_{\rm R}^4 H^2 }{2 a^2  k^2} +\frac{a_{\rm R}^4  H^4+ a_{\rm R}^4 H^4}{2k^4}\right) \sin kr \\ \nonumber
	&=& \lim_{r \to 0}  \left[ \frac{1}{4 \pi^2 a^4}  \frac{1}{r}   \int_0^{\infty} dk  k^2 \sin kr +\frac{a_{\rm R}^4 H^2}{8 \pi^2 a^6}  \frac{1}{r}   \int_0^{\infty} dk  \sin kr \right. \\  \label{rholimRD2}
	&& \left.  + \frac{1}{4 \pi^2 a^4}  \frac{1}{r}   \int_0^{\infty} dk \left( \frac{a_{\rm R}^4  H^4+ a_{\rm I}^4 H^4}{2k^2} \right) \sin kr   \right] .
\end{eqnarray}
Changing integration variables to $q=kr$, one can identify the terms that survive in the $r\to 0$ limit, given by 
\eqn{rholimRD3}{\begin{aligned}
	\rho_{\rm div}  &= \lim_{r \to 0}  \left[ \frac{1}{4 \pi^2 a^4}  \int_0^{\infty} dq \sin q  \left(\frac{1}{r^4}  q^2+\frac{a_{\rm R}^4 H^2}{2  a^2}  \frac{1}{r^2}  + \frac{a_{\rm R}^4 H^4}{2q^2}\left(1 + \frac{a^4_{\rm I}}{a^4_{\rm R}}\right)  \right)  \right]\\ 
	&= \lim_{r \to 0}  \left[ -\frac{2}{4 \pi^2 a^4}  \frac{1}{r^4} + \frac{a_{\rm R}^4 H^2}{8 \pi^2 a^6}  \frac{1}{r^2} + \frac{a_{\rm R}^4 H^4}{8 \pi^2 a^4}\left(1 + \frac{a^4_{\rm I}}{a^4_{\rm R}}\right)\left[\gamma_E - 1 + \log kr \right] \right]\\ &= \lim_{\sigma \to 0}  \left[ -\frac{2}{4 \pi^2 \sigma^4} + \frac{1}{8 \pi^2 \sigma^2}\frac{H^2}{(a/a_{\rm R})^4} + \frac{H^4\left(1 + e^{-4\mathcal N_{\rm tot}}\right)}{8 \pi^2 (a/a_{\rm R})^4}\left[\gamma_E - 1 + \log k_p\sigma \right] \right].
\end{aligned}}
We note that the power law divergences identified using physical momentum cutoffs in Eqs. \ref{hcemt1}, \ref{hcemt2} can be read off from above up to the expected scheme difference in the coefficients. However, as advertised, the logarithmic divergence is reproduced with the same coefficients as those obtained with a physical momentum cutoff Eq. \ref{corho}, or via dimensional regularization Eq. \ref{rholog}. 

\section{\label{app:gw} Improved stress tensor for gravitational waves}

The Isaacson form of the stress tensor is defined as the averaged high frequency part of the second order expansion of Einstein equations (\cite{Isaacson} in the conventions and notation of \cite{Maggiore})
\begin{equation}\label{Isaacson def Tmunu}
	T_{\mu \nu}{ }^{\rm gw, hf}  :=  -\frac{1}{8 \pi  G_N}  \left\langle  \delta^2R_{\mu \nu}-\frac{1}{2} g_{\mu \nu} g_{\alpha \beta} \delta^2 R^{\alpha \beta} \right\rangle,
\end{equation}
where $g_{\mu \nu} $ is the background metric (defined as $\tilde{g}_{\mu \nu} =g_{\mu \nu} +h_{\mu \nu} $) and $\delta^2R_{\mu \nu}$ is the second order perturbation of the Ricci tensor, and therefore quadratic in $h_{\mu \nu}$. The averaging scheme used in \cite{Isaacson} is the Brill-Hartle (BH) averaging scheme \cite{Brill:1964zz} which as a corollary allows one to neglect divergences, freely integrate by parts within expectation values, and to commute covariant derivatives (thus implicitly assuming that the curvature scale of the background is much smaller than the frequencies of interest). After the BH averaging (and fixing the gauge as $D_{\rho}h^{\rho \nu}=0 $, $h_{\mu}{}^{ \mu}=h=0$), the resulting stress energy tensor is given by
\begin{equation}\label{BH averaged Tmunu}
	T_{\mu \nu}{ }^{\rm gw,hf} = \frac{1}{32 \pi G_{N}}\langle \nabla_\mu h_{\rho \sigma} \nabla_\nu h^{\rho \sigma} \rangle_{\rm BH}.
\end{equation}
Consequently, the energy density for a stochastic background of gravitational waves is found to be
\begin{equation}\label{Rho gw literature 0}
	\begin{aligned}
		\rho_{\rm gw}^{\rm Isc} = \frac{1}{32 \pi a^2 G_N} \delta^{i m} \delta^{j \ell}\left\langle h^{\prime}_{i j}(\tau, k) h^{\prime}_{l m}(\tau, k)\right\rangle,
	\end{aligned}
\end{equation}
where Eq. \ref{Rho gw literature 0} is valid only for sufficiently high frequency signals such that the curvature of the background can be neglected. Therefore, it behooves us to re-examine the derivation of the stress tensor for gravitational waves without resorting to approximations that may be at odds with the need to incorporate wavelengths beyond this approximation in any intermediate steps in following through the process of renormalization on cosmological spacetimes. In the following we re-derive Eq. \ref{Rho gw literature 0} without the assumptions of having an high frequency signal propagating on a flat background.

We compute the stress energy tensor of gravitational waves $T^{\rm gw}_{\mu \nu}$ following \cite{Isaacson}  and  \cite{Maggiore} where the stress energy tensor of gravitational waves is defined as the averaged second order perturbed Einstein equations\footnote{One can also independently derive what follows by expanding the action to second order in perturbations and varying with respect to the background metric, taking care to properly address gauge fixing and ghost terms that are necessary in the context of quantum expectation values, the net result of which will be Eq \ref{eq:Tmunu 3} and \ref{rhoin} \cite{NP}.}

\begin{eqnarray} \nonumber
		T^{\rm gw}{}_{\mu}{}^{ \nu}& := & -\frac{1}{8 \pi G_N} \left\langle \delta^2 G_{\mu}^{ \nu}\right\rangle \\ \nonumber
		&=&-\frac{1}{8 \pi G_N} \left\langle \delta^2g^{\nu \alpha} R_{\mu \alpha} +\delta^1g^{\nu \alpha} \delta^1R_{\mu \alpha} + g^{\nu \alpha} \delta^2R_{\mu \alpha}  \right. \\ \label{eq:Tmunu 1}
		&& \left. -\frac{1}{2} \delta_{\mu}^{ \nu} \left( \delta^2g^{\alpha \beta} R_{ \alpha \beta}  +\delta^1g^{\alpha \beta} \delta^1R_{ \alpha \beta}+g^{\alpha \beta}  \delta^2R_{ \alpha \beta}\right)\right\rangle .
\end{eqnarray}
In the equation above, $g_{\mu \nu}$ is the Friedman metric in the mostly plus convention in conformal time and (see \cite{Maggiore} for details)
\begin{eqnarray} \nonumber
		&&\delta^1g_{\mu \nu}=h_{\mu \nu} \quad \quad \delta^1g^{\mu \nu}=-h^{\mu \nu} \quad \quad 	\delta^2g^{\mu \nu}=h^{\mu \alpha}h_{\alpha}^{ \nu}\\ \nonumber
	&&\delta^1R_{\mu \nu}=-\frac{1}{2} \Box h_{\mu \nu}-\frac{1}{2} D_{\nu} D_{\mu}h+\frac{1}{2}D^{\rho} D_{\mu}h_{ \nu \rho}+\frac{1}{2}D^{\rho} D_{\nu}h_{ \mu \rho}\\ \nonumber
	&&\delta^2R_{\mu \nu}=\frac{1}{2} g^{\rho \sigma} g^{\alpha \beta}\left[\frac{1}{2} D_{\mu} h_{\rho \alpha} D_{\nu} h_{\sigma \beta}+\left(D_{\rho} h_{\nu \alpha}\right)\left(D_{\sigma} h_{\mu \beta}-D_{\beta} h_{\mu \sigma}\right.\right)\\ \nonumber
	&&\left.\quad \quad \quad +h_{\rho \alpha}\left(D_{\nu} D_{\mu} h_{\sigma \beta}+D_{\beta} D_{\sigma} h_{\mu \nu}-D_{\beta} D_{\nu} h_{\mu \sigma}-D_{\beta} D_{\mu} h_{\nu \sigma}\right)\right.\\ \label{eq:perturbations metric}
	&&\left.\quad \quad \quad +\left(\frac{1}{2} D_{\alpha} h_{\rho \sigma}-D_{\rho} h_{\alpha \sigma}\right)\left(D_{\nu} h_{\mu \beta}+D_{\mu} h_{\nu \beta}-D_{\beta} h_{\mu \nu}\right)\right].
	\end{eqnarray}
Using the definitions above and fixing the gauge ($D_{\rho}h^{\rho \nu}=0 $, $h_{\mu}{}^{ \mu}=h=0$) the stress energy tensor in (\ref{eq:Tmunu 1}) results
\begin{equation}\label{eq:Tmunu 2}
	\begin{aligned}
		T^{\rm gw}{}_{\mu}{}^{ \nu}=&-\frac{1}{8 \pi G_N}\left\langle-\frac{1}{2}h^{\nu}{}_{ \alpha}h_{\mu}{}^{ \sigma}R^{\alpha}{}_{ \sigma}+\frac{1}{2}h^{\rho}{}_{ \alpha}h_{\rho}{}^{ \nu}R^{\alpha}{}_{ \mu}+h_{\alpha}{}^{ \nu}h_{\sigma}{}^{ \rho}R^{\sigma \alpha}{}_{\rho \mu}+\frac{1}{2} h^{\nu}{}_{ \alpha}\Box h^{\alpha}{}_{ \mu} \right.\\
		&\left. +\frac{1}{4}D_{\mu}h^{\beta}{}_{ \alpha}D^{\nu}h_{\beta}{}^{ \alpha} +\frac{1}{2}D_{\sigma}h^{\nu}{}_{ \alpha}D^{\sigma}h_{\mu}{}^{ \alpha}-\frac{1}{2}D_{\sigma}h^{\nu}{}_{ \alpha}D^{\alpha}h_{\mu}{}^{ \sigma}+\frac{1}{2}h^{\sigma}{}_{ \alpha}D_{\mu}D^{\nu}h_{\sigma}{}^{ \alpha}\right.\\
		&\left.+\frac{1}{2}h^{\sigma}{}_{ \alpha}D^{\alpha}D_{\sigma}h_{\mu}{}^{ \nu}-\frac{1}{2}h^{\sigma}{}_{ \alpha}D^{\alpha}D_{\mu}h_{\sigma}{}^{ \nu}-\frac{1}{2}h^{\alpha}{}_{ \sigma}D_{\alpha}D^{\nu}h_{\mu}{}^{ \sigma}\right.\\
		&\left.-\frac{1}{2}\delta^{\nu}{}_{ \mu} \left( h_{\beta}{}^{ \alpha}h_{\sigma}{}^{ \rho}R^{\sigma \beta}{}_{\rho \alpha}+h^{\beta}{}_{ \alpha}\Box h^{\alpha}{}_{ \beta}+\frac{3}{4}D_{\rho}h^{\alpha}{}_{ \beta}D^{\rho}h_{\alpha}{}^{ \beta}-\frac{1}{2}D_{\sigma}h^{\beta}{}_{ \alpha}D^{\alpha}h_{\beta}{}^{ \sigma} \right)\right\rangle .
	\end{aligned}
\end{equation}
At this stage, Brill-Hartle averaging would result in the Maccallum-Taub averaged stress tensor \cite{Maccallum:1973gf}, which upon further integrations by parts within the spatial and temporal averaged integrals in addition to commuting covariant derivatives would bring the latter into the Isaacson form Eq. \ref{BH averaged Tmunu}. Instead, we persist with Eq. \ref{eq:Tmunu 2} as it is. To find the energy density of gravitational waves $\rho_{\rm gw}$, we have to specify the covariant derivatives in Eq \ref{eq:Tmunu 2}. Using that in conformal time the only non-vanishing Christoffel symbols are $\Gamma_{00}^{ 0}=\frac{a^{\prime}}{a}$, $\Gamma_{0j}^{ i}=\frac{a^{\prime}}{a} \delta_{j}{}^{ i}$ and $\Gamma_{ij}^{ 0}=\frac{a^{\prime}}{a}\delta_{ij}$, and considering only the transverse traceless part of the metric as the propagating degrees of freedom\footnote{In this derivation we refer to the scalar-vector-tensor decomposition and the fact that, after fixing of the residual gauge, only the transverse traceless part of the metric $h_{ij}$ are the propagating degrees of freedom.}, $T_{0}{}^{ 0}$ results
\begin{equation}\label{eq:Tmunu 3}
	\begin{aligned}
		T_{0}{}^{ 0}=&\frac{-1}{8 \pi a^2 G_N}\left\langle \frac{1}{8} h^\prime_{j}{}^{ i} h^\prime_{i}{}^{  j}- \frac{3}{8} \partial_k h_{j}{}^{ i} \partial^k h_{i}{}^{ j} - \frac{1}{2}  h_{j}{}^{ i} \partial_k\partial^k h_{i}{}^{ j}+  \mathcal{H}h_{j}{}^{ i} h^\prime_{i}{}^{ j}+\frac{1}{4} \partial_k h_{j}{}^{ i} \partial^j h_{i}{}^{ k} \right\rangle .
	\end{aligned}
\end{equation}
Consequently, in the coincidence limit ($\rho_{\rm gw} :=  \lim_{y\to x}\rho_{\rm gw}(\tau;x,y)$) the energy density of gravitational waves can be expressed as
\begin{equation}\label{rhoin}
	\rho_{\rm gw}=\frac{1}{64 \pi  a^2 G_N}\left\langle  h^\prime_{j}{}^{ i} h^\prime_{i}{}^{  j} -3 \partial_k h_{ij} \partial^k h^{ ij} - 4 h_{ij} \partial_k\partial^k h^{ ij}+2\partial_k h_{ij} \partial^j h^{i k} + 8 \mathcal{H}h_{j}{}^{ i} h^\prime_{i}{}^{ j}\right\rangle .
\end{equation}
In spite of not invoking any additional averaging prescriptions, the expectation value featuring in the improved stress tensor above is still doing a lot of heavy lifting. We stress that this is purely a quantum expectation value (strictly, an in-in correlation function) at some fixed time with no extra spatial or temporal averaging invoked. Furthermore, the expectation value presumes a density matrix, which we take to be Bunch Davies vacuum for questions of vacuum tensor perturbations as discussed in the main body of the draft. A far more interesting story beyond the scope of the present commences once one incorporates the effects non-trivial density matrices, and interactions induced by gravitational non-linearities, as well as those induced by matter couplings. 

\end{appendices}

\bibliography{sn-bibliography}

\section*{Statements and Declarations}

\textbf{Funding}  \\ 
Author A. Negro was supported by the Casimir PhD Travel Grant to visit the Perimeter Institute, where some of the work leading to this review was initiated.\\
\textbf{Competing interests } \\
The authors have no relevant financial or non-financial interests to disclose. \\
\textbf{Author Contributions} \\
Equal contribution. \\
\textbf{Data Availability}  \\
Non applicable.

\end{document}